\documentclass[twocolumn]{aastex7} 
\usepackage{amsmath, mathtools, amsthm, amssymb}
\usepackage{enumitem}
\usepackage[toc,page]{appendix}
\hypersetup{colorlinks=true, allcolors=cyan}
\usepackage{soul}
\graphicspath{ {./} }
\usepackage{xcolor}
\usepackage{nameref}
\usepackage[super]{nth}
\usepackage{booktabs}
\usepackage{bm}

\newcommand{\msol}{\ensuremath{\rm{M_{\odot}}}}
\newcommand{\NC}{\textsc{NewCluster}}
\newcommand{\HM}{\textsc{AdaptaHOP}}
\newcommand{\rtwo}{\ensuremath{{R_{\rm 200}}}}
\newcommand{\mtwo}{\ensuremath{{M_{\rm 200}}}}
\newcommand{\rvir}{\ensuremath{{R_{\rm vir}}}}
\newcommand{\mvir}{\ensuremath{{M_{\rm vir}}}}
\newcommand{\rmax}{\ensuremath{{R_{\rm max}}}}
\newcommand{\tinf}{\ensuremath{{t_{\rm inf}}}}
\newcommand{\gid}[1]{\texttt{(#1)}}
\newcommand{\yfrom}{\texttt{FIRST}}
\newcommand{\yhead}{\texttt{HEAD}}
\newcommand{\ytail}{\texttt{TAIL}}
\newcommand{\ylast}{\texttt{FINAL}}
\newcommand{\afe}{[$\alpha$/Fe]}
\newcommand{\feh}{[Fe/H]}

\newcommand{\zing}{0.56}
\newcommand{\ting}{Dec. 2025}
\newcommand{\bcgicl}{\textsf{BCG+ICL}}
\newcommand{\iclA}{\textsf{stripped}}
\newcommand{\iclB}{\textsf{disrupted}}
\newcommand{\iclC}{\textsf{in-situ}}
\newcommand{\iclD}{\textsf{preprocessed}}

\usepackage[dvipsnames]{xcolor}

\sethlcolor{Lavender} 

\newcommand{\eadd}[1]{#1} 

\newif\ifshowedel
\showedeltrue   
\ifshowedel
  \newcommand\edel[1]{\bgroup\markoverwith{\textcolor{violet}{\rule[0.5ex]{2pt}{0.8pt}}}\ULon{#1}}
\else
  \newcommand\edel[1]{} 
\fi

\shorttitle{IntraNewCluster Light}
\shortauthors{Seyoung Jeon et al.}

\begin{document}

\title{On the Origin of Intracluster Light based on the High-resolution Simulation, \NC}

\author[0000-0002-1270-4465]{Seyoung Jeon}\email{syj3514@yonsei.ac.kr}
\affil{Department of Astronomy and Yonsei University Observatory, Yonsei University, 50 Yonsei-ro, Seodaemun-gu, Seoul 03722, Republic of Korea}

\author[0000-0002-2873-8598]{Emanuele Contini}\email{emanuele.contini82@gmail.com}
\affil{Department of Astronomy and Yonsei University Observatory, Yonsei University, 50 Yonsei-ro, Seodaemun-gu, Seoul 03722, Republic of Korea}

\author[0000-0001-9939-713X]{San Han}\email{san.han@iap.fr}
\affil{Institut d’Astrophysique de Paris, CNRS and Sorbonne Université, UMR 7095, 98 bis Boulevard Arago, F-75014 Paris, France}

\author[0000-0002-0184-9589]{Jinsu Rhee}\email{jinsu.rhee@gmail.com}
\affil{Institut d’Astrophysique de Paris, CNRS and Sorbonne Université, UMR 7095, 98 bis Boulevard Arago, F-75014 Paris, France}

\author[0000-0003-2939-8668]{Garreth Martin}\email{Garreth.Martin@nottingham.ac.uk}
\affil{School of Physics and Astronomy, University of Nottingham, University Park, Nottingham NG7 2RD, UK}

\author[0000-0002-4391-2275]{Juhan Kim}\email{kjhan0606@gmail.com}
\affil{Center for Advanced Computation, Korea Institute for Advanced Study, 85 Hoegi-ro, Dongdaemun-gu, Seoul 02455, Republic of Korea}

\author[0000-0002-6810-1778]{Jaehyun Lee}\email{jaehyun@kasi.re.kr}
\affiliation{Korea Astronomy and Space Science Institute, 776, Daedeokdae-ro, Yuseong-gu, Daejeon 34055, Republic of Korea}

\author[0000-0002-3950-3997]{Taysun Kimm}\email{tkimm@yonsei.ac.kr}
\affil{Department of Astronomy and Yonsei University Observatory, Yonsei University, 50 Yonsei-ro, Seodaemun-gu, Seoul 03722, Republic of Korea}

\author[0000-0003-0695-6735]{Christophe Pichon}\email{pichon@iap.fr}
\affil{Institut d’Astrophysique de Paris, CNRS and Sorbonne Université, UMR 7095, 98 bis Boulevard Arago, F-75014 Paris, France}
\affil{Kyung Hee University, Dept. of Astronomy \& Space Science, Yongin-shi, Gyeonggi-do 17104, Republic of Korea}

\author[0009-0007-8611-3813]{Gyeong-Hwan Byun}\email{byunastro@yonsei.ac.kr}
\affil{Department of Astronomy and Yonsei University Observatory, Yonsei University, 50 Yonsei-ro, Seodaemun-gu, Seoul 03722, Republic of Korea}

\author[0000-0003-0225-6387]{Yohan Dubois}\email{dubois@iap.fr}
\affil{Institut d’Astrophysique de Paris, CNRS and Sorbonne Université, UMR 7095, 98 bis Boulevard Arago, F-75014 Paris, France}

\author[0000-0003-2285-0332]{Corentin Cadiou}\email{corentin.cadiou@iap.fr}
\affil{Institut d’Astrophysique de Paris, CNRS and Sorbonne Université, UMR 7095, 98 bis Boulevard Arago, F-75014 Paris, France}

\author[0000-0002-0858-5264]{J. K. Jang}\email{starbrown816@yonsei.ac.kr}
\affil{Department of Astronomy and Yonsei University Observatory, Yonsei University, 50 Yonsei-ro, Seodaemun-gu, Seoul 03722, Republic of Korea}

\author[0000-0002-4556-2619]{Sukyoung K. Yi}\email{yi@yonsei.ac.kr}
\affil{Department of Astronomy and Yonsei University Observatory, Yonsei University, 50 Yonsei-ro, Seodaemun-gu, Seoul 03722, Republic of Korea}

\begin{abstract}

Intracluster light (ICL) is a key component of galaxy clusters, with the potential to trace their dynamical assembly histories and the underlying dark matter distribution.
Despite these prospects, its faint nature makes a consensus on its origin or population properties difficult to achieve, both in observations and simulations.
In the hope of finding a breakthrough, we utilize the ongoing high-resolution cluster simulation, \NC.
By classifying billions of particles in and around the cluster with a rigorous tracking procedure, we find that the majority of the ICL originates from satellites, including surviving and disrupted galaxies.
Another notable finding is that the \textit{preprocessed} component follows the density profile of dark matter better than the other components and has distinctive properties: old age, low metallicity, and enhanced $\alpha$-element abundance.
We further investigate the orbital dynamics, and our results demonstrate that the stripped fraction of satellites is primarily determined by the \textit{time since infall} and the pericenter distance.
By linking the demographic, chemical, and orbital properties of ICL stars to their origins, this work proposes a quantitative approach for tracing the assembly history of galaxy clusters from the ICL.

\end{abstract}


\section[]{Introduction}\label{sec:1intro}

Intracluster light (ICL) is a diffuse stellar component residing in the gravitational potential of a galaxy cluster, without being bound to the potential of individual satellite galaxies \citep{Mihos2016, Montes2019, Contini2021, Montes2022a}.
Since its first detection by \citet{Zwicky1937, Zwicky1951} in the Coma cluster, it has been recognized as an important cluster component and visible evidence of the hierarchical structure formation via numerous mergers and tidal stripping \citep{Willman2004, Mihos2005, Zibetti2005, Monaco2006, Murante2007, Contini2014, Montes2022a, Brough2024, JimenezTeja2025}.
Therefore, studying the ICL is anticipated to reveal the past assembly history of galaxy clusters.
Furthermore, the ICL is expected to follow the cluster's total dynamical mass distribution \citep{Montes2019, ContrerasSantos2024, Yoo2025}, which is dominated by dark matter.
As a consequence, its observation can unlock an independent way to trace dark matter, complementary to X-ray emission \citep{Borgani2001} or gravitational lensing analysis \citep{Kochanek2006, Clowe2006, Cha2023}.

Despite these promising prospects, the depth of analysis is relatively limited due to its faint and diffuse nature.
The typical surface brightness of the ICL is fainter than $\sim26\,{\rm mag\,arcsec^{-2}}$ in the optical range, and can reach $\sim30\,{\rm mag\,arcsec^{-2}}$ or beyond, which requires substantial sensitivity and resolution \citep{Mihos2005, Zibetti2005, Rudick2011}.
In recent decades, modern deep imaging surveys have reached this limit \citep{Trujillo2016, Montes2021, Gonzalez2021, Montes2022b, MartinezLombilla2023, Joo2023, Kluge2025}, enabling statistical studies.
However, the uncertainty in detection is still huge, and many properties obtained by different studies have not yet converged.
For example, the primary property of the ICL is the mass fraction relative to the total stellar mass or the brightest cluster galaxy (BCG) mass in a galaxy cluster.
Current studies suggest the fraction measured is 5--50 percent \citep[][see references therein]{Contini2024book}, but it is still unclear whether this large range of fractions comes from cluster-to-cluster variation or methodological differences \citep{Rudick2011, Kimmig2025}.
Moreover, how the fraction changes with redshifts or halo properties (e.g., mass, concentration, or dynamical states) is more complex \citep{GoldenMarx2025}, or even whether it shows little variation, remains a debated topic \citep{Ko2018, Joo2023, Contini2024a}.

Another important issue is the origin of the ICL.
Tidal stripping of satellite galaxies, by-products of mergers with the BCG, or preprocessing are considered important channels \citep{Mihos2003, Chun2024, Contini2024a}; however, which process has a dominant role or how these different origins affect the stellar populations are still under debate.
Perhaps the most difficult issue is the absence of consensus on definitions.
Specifically, as recently introduced in \citet{Brough2024}, there are various ways of determining the ICL itself, such as using a surface brightness threshold \citep{Montes2014, MartinezLombilla2023}, masking galactic light \citep{Annunziatella2016, Ahad2023, Bellhouse2025}, or a wavelet technique \citep{Adami2005, Ellien2021}.
This lack of agreement fundamentally stems from the nontrivial ICL detection.

Numerical simulations, in parallel, have provided a powerful tool to study the formation of the ICL.
Early on, \citet{Murante2004} identified the diffuse light in the simulated galaxy cluster and found that the diffuse light is more centrally concentrated and older than the galaxy light.
Using the TNG300 simulation \citep{Nelson2019}, \citet{MontenegroTaborda2023} showed that 90\% of the ICL mass is contributed by ``accreted'' stars.
Recently, \citet{Kimmig2025} compared several independent cosmological simulations, including \textsc{Magneticum Pathfinder} \citep{Dolag2025}, \textsc{TNG100} \citep{Nelson2019}, \textsc{Horizon-AGN} \citep{Dubois2014bHAGN}, and \textsc{Hydrangea} \citep{Bahe2017}; they found that the formation redshift of galaxy clusters is tightly correlated with the mass fraction of the sum of BCG and ICL (hereafter, \bcgicl) relative to the cluster's total stellar mass.
Complementing these hydrodynamical efforts, semi-analytic models (SAMs) have also provided key insights.
For example, since \citet{Monaco2006} predicted a total amount of diffuse stellar components with the \textsc{MORGANA} model, \citet{Contini2024c} introduced the \textsc{FEGA} model that provides a large statistical sample of clusters, a task that remains computationally too expensive for fully-hydrodynamical simulations.
Despite the merits of using simulations, the faint nature of the ICL still remains a major challenge for theoretical studies as well.
A galaxy cluster is a very massive object, and it requires much more computational resources than field environments to maintain particle mass and spatial resolutions.
Given the limited resolution of conventional cosmological simulations, their resolution has been insufficient to fully resolve the detailed structure and stellar populations of faint tidal features and merger relics \citep{Martin2024, Brown2024, Kimmig2025}.
Although the semi-analytic approach gives us significant benefits, particularly in statistical trends \citep{Guo2011, Contini2023, Contini2024b}, it is also inherently limited by its nature of solving analytic equations for galaxy properties rather than employing individual stellar particles.
For a detailed analysis of the ICL, it is thus important to resolve the dynamics of stars finely to track the dynamical stripping of stars.

In response to this demand, we use \NC, a high-resolution cosmological hydrodynamical simulation targeting a galaxy cluster \citep{Han2025b_ncintro}.
As we will describe the details of \NC\ in Section~\ref{sec:2method_1sim}, we hereby highlight the strengths of \NC\ in studying the ICL.
\NC\ achieves substantially improved resolutions compared to other cluster-scale simulations \citep[cf.][Table 1]{Tremmel2019}, specifically a stellar mass resolution of $2\times10^4\,\msol$ and a spatial resolution of $\rm 68\,pc$, allowing studies down to a limiting $r$-band surface brightness of $\rm 33.5\, mag\,arcsec^{-2}$ at $z=0.8$.
We also have an order of magnitude higher output cadence of $\sim15\,\mathrm{Myr}$ (i.e., $\sim1000$ snapshots for redshifts 50 to 0) than the conventional setting of $\sim 100$ snapshots.
This enables better tracking of galaxies across snapshots and capturing the orbital motion in dense environments.

Throughout this work, utilizing \NC, we study the origin of the ICL from the detailed orbital history of individual stellar particles, which has never been attempted at this resolution.
In Section~\ref{sec:2methods}, we describe information on \NC, the halo/galaxy identification, the merger tree construction, and the definition of satellite galaxies.
In Section~\ref{sec:3class}, we describe comprehensively the procedure that defines \bcgicl\ stars and further divides them into subsamples by their origins.
The overall results and their demographics are presented in Section~\ref{sec:4demographics}.
Section~\ref{sec:5population} presents the analysis of the properties of their stellar populations.
In Section~\ref{sec:6orbit}, we further investigate the role of the orbital motion in the formation of the ICL.
We then discuss the current limitations and the future outlook in Section~\ref{sec:7limit}, and finally, we summarize our findings and conclude in Section~\ref{sec:8conclusion}.

\section{Methods}\label{sec:2methods}

In this section, we describe the details of the simulation.
Building merger trees and classifying satellites in the galaxy cluster are critical factors of this work, so we give particular attention to them.

\subsection{Numerical simulations}\label{sec:2method_1sim}

%
\begin{figure}[htb!]
    \centering
    \includegraphics[width=0.45\textwidth]{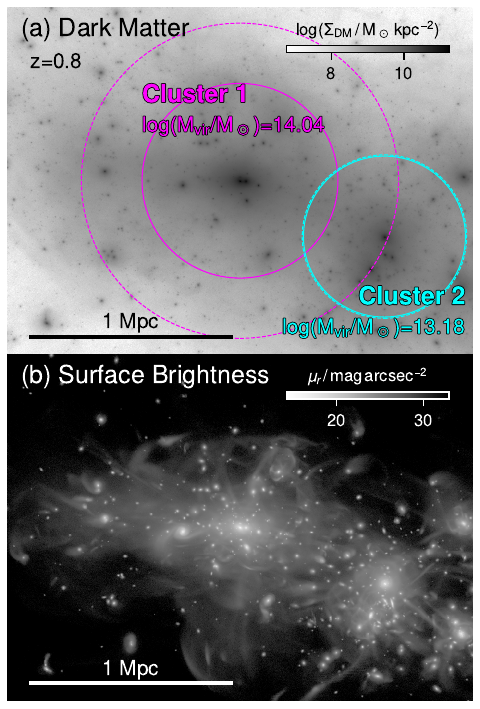}
    \caption{
         Overview of the \NC\ simulation centered on the primary cluster halo at $z=0.79$.
         Panel~(a) presents a dark matter density map (grayscale), with the virial radii of the two cluster halos (magenta and cyan circles).
         The dotted line shows the virial radius (\rvir) from \HM, and the dashed line is the \rtwo, which are described in Section~\ref{sec:2method_2halo}.
         Panel~(b) shows the corresponding $r$-band surface brightness map derived from the stellar density.
         Note that faint but complex tidal features are resolved in the surface brightness map.
        }
    \label{fig01_overall}
\end{figure}
%

\NC\ is a cosmological zoom-in simulation targeting a galaxy cluster with a virial mass comparable to that of the Virgo cluster at $z=0$ \citep{Han2025b_ncintro}.
The simulation is conducted using the RAMSES-yOMP code \citep{Han2025a_yomp}, the hybrid parallelized version of the adaptive mesh refinement (AMR) code, RAMSES \citep{Teyssier2002}, assuming the \citet{Chabrier2003} initial mass function and the WMAP7 cosmology \citep{Komatsu2011}.
The initial volume of the targeted zoom-in region is $\sim23,300\,\textrm{Mpc}^3$ at the center of the total $(100\,h^{-1}\,\textrm{Mpc})^3$ box.
In this zoom-in region, each cell can form stars following the gravo-thermo-turbulent model \citep{Kimm2017} when its density exceeds $5\,\textrm{H\,cm}^{-3}$.
The subsequent stellar feedback process is composed of Type II and Ia supernovae with prescriptions from \citet{Kimm2014}.
\NC\ also tracks the evolution of chemical elements (H, D, He,
C, N, O, Mg, Fe, Si, and S) ejected from stellar wind \citep{Schaller1992, Maeder2000}, Type Ia supernovae \citep{Kobayashi2006}, and Type II supernovae \citep{Iwamoto1999}.
The dual-mode feedback (i.e., radio and quasar modes by the accretion rates) from active galactic nuclei (AGN) is also embedded in \NC \citep{Dubois2014a, Dubois2014bHAGN}.
For other detailed prescriptions, we refer readers to the introductory paper of \NC\ \citep{Han2025b_ncintro}.

In Figure~\ref{fig01_overall}, we present an overview of \NC\ at $z=0.79$.
Panel~(a) shows the dark matter density map and the primary and secondary halos.
As the two halos are about to merge, the primary halo is expected to be highly perturbed in the near future.
To avoid such unrelaxed dynamical states, we limit our study to the snapshot at $z=0.79$.
This snapshot is also the latest one that has been fully post-processed for analysis.\footnote{\NC\ is still ongoing and has currently passed $z\sim\zing$ as of \ting.}

\NC\ has many updated astrophysical prescriptions compared to its predecessors---\textsc{Horizon-AGN} \citep{Dubois2014a} and \textsc{NewHorizon} \citep{Dubois2021}---, such as chemical elements, on-the-fly dust evolution, Lagrangian gas tracer particles \citep{Genel2013, Cadiou2019}, or dynamical friction of a supermassive black hole \citep{Dubois2014a, Pfister2019, Han2025b_ncintro}.
However, the most important strength related to this work is the resolution.
The spatial resolution is $\rm 68\,pc$ in the most refined cell, and the stellar mass resolution is $\rm \sim 2 \times 10^4\,\msol$.
This allows us to resolve the primary cluster region into $\sim\,200$ million stellar particles, which is an unprecedented scale.
Low surface brightness features, including the ICL, require fine resolutions to better reproduce their diffuse nature and complicated dynamics.
Using a suite of TNG simulations, \citet{Lovell2025} assessed the numerical effects and demonstrated that the resolution highly affects the stripped mass of satellites.
Although they argued that the stripping times of the dark matter are converged in their flagship runs---TNG50-1 ($m_{\rm DM}=4\times10^5\,\msol$ and $\epsilon_{\rm DM}=0.29\,\mathrm{kpc}$) to TNG300-1 ($m_{\rm DM}=5.9\times10^7\,\msol$ and $\epsilon_{\rm DM}=1.48\,\mathrm{kpc}$), where $m_{\rm DM}$ and $\epsilon_{\rm DM}$ are the dark matter particle mass resolution and softening length, respectively---, the stripped mass and stripping times of stellar particles are not converged in their resolutions.
This magnifies the strength of our spatial resolution of $68\,\mathrm{pc}$ in minimizing the numerical artifacts on resolving tidal features.
For example, in Figure~\ref{fig01_overall}(b), we present the $r$-band surface brightness map of \NC.
The faint but highly disturbed features are prominent in the surface brightness map.

\subsection{Halo (galaxy) identification}\label{sec:2method_2halo}
We utilize the \HM\ halo-finding algorithm to identify galaxies (\textsc{GalaxyMaker}) and dark matter halos (\textsc{HaloMaker}) from stellar and dark matter particles, respectively.
As a detailed description of the algorithm can be found in \citet{Aubert2004, Tweed2009}, we briefly outline the main procedure here.
This method finds local density maxima and saddle points that lie above a minimum density threshold---conventionally $80\times\rho_{\rm crit}$ for halos and $178\times\rho_{\rm crit}$ for galaxies, where $\rho_{\rm crit}$ is the critical density of the Universe---, which are regarded as initial guesses of the center and boundary of structures.
After identifying the topological structures, \HM\ iteratively eliminates structures that do not physically satisfy the input conditions, such as the density threshold or the gradient of the substructure density.
It also builds the structure-substructure hierarchy based on spatial occupation.
For each ground-level structure, the structures hierarchically connected to the most massive substructure are integrated into a single structure, which is designated as the main structure, while the remaining structures are tagged as substructures (i.e., subhalos or satellite galaxies).
To robustly identify the center, \HM\ iteratively shrinks concentric shells toward the densest region rather than simply selecting the densest particle or the center of mass.
In this work, we use the term \rmax, which is the maximum distance from the galaxy/halo center to any of its member particles, indicating the spatial occupation of the structure.

With these structures, \HM\ calculates the physical properties, such as the virial radius, the virial mass, or the spin parameters.
For the virial radius of a dark matter halo, \HM\ directly calculates the virial condition using the member dark matter particles:
\begin{equation}
    \eta_{\rm vir} = \left|\frac{\,2T+W\,}{\,T+W\,}\right|,
\end{equation}
where $\eta_{\rm vir}$ is the virial ratio between the virial residual and the total energy, $T$ is the kinetic energy, and $W$ is the potential energy.
By iteratively shrinking the concentric ellipsoidal radius from where $T+W<0$ and computing $\eta_{\rm vir}$ at each step, the virial radius (\rvir) is defined at the radius where $\eta_{\rm vir}$ first drops below $0.2$.\footnote{This threshold is a default setting in AdaptaHOP, which is a small and non-zero threshold to account for minor deviations from a perfectly relaxed state.}
If a halo is too unrelaxed for this criterion to be met---i.e., $\eta_{\rm vir}$ never falls below $0.2$ over the inner sweep---, \HM\ regards the radius within which the density is greater than $\Delta_{\rm vir}(z)\times \rho_{\rm mean}$ as the virial radius, where $\Delta_{\rm vir}(z)$ is the virial overdensity relative to the mean matter density computed from a spherical top-hat collapse calculation, and $\rho_{\rm mean}$ is the mean density of the Universe.
The virial mass (\mvir) is defined as the dark matter mass within an ellipsoid corresponding to \rvir.
However, because this calculation considers only the member dark matter particles, it can be offset from the total dynamical mass.
Therefore, we further calculate \rtwo\ as the radius within which the mean enclosed density is 200 times the critical density, taking baryon components into account, and the corresponding mass is \mtwo.
In this work, we adopt \rtwo\ and \mtwo\ as the halo virial radius and mass.
For reference, the \mvir\ and \mtwo\ of the primary halo are $1.09\times10^{14}\,\msol$ and $7.76\times10^{13}\,\msol$, and those of the secondary halo are $1.52\times10^{13}\,\msol$ and $4.38\times10^{13}\,\msol$ at $z=0.79$.

%
\begin{figure*}[htb]
    \centering
    \includegraphics[width=0.80\textwidth]{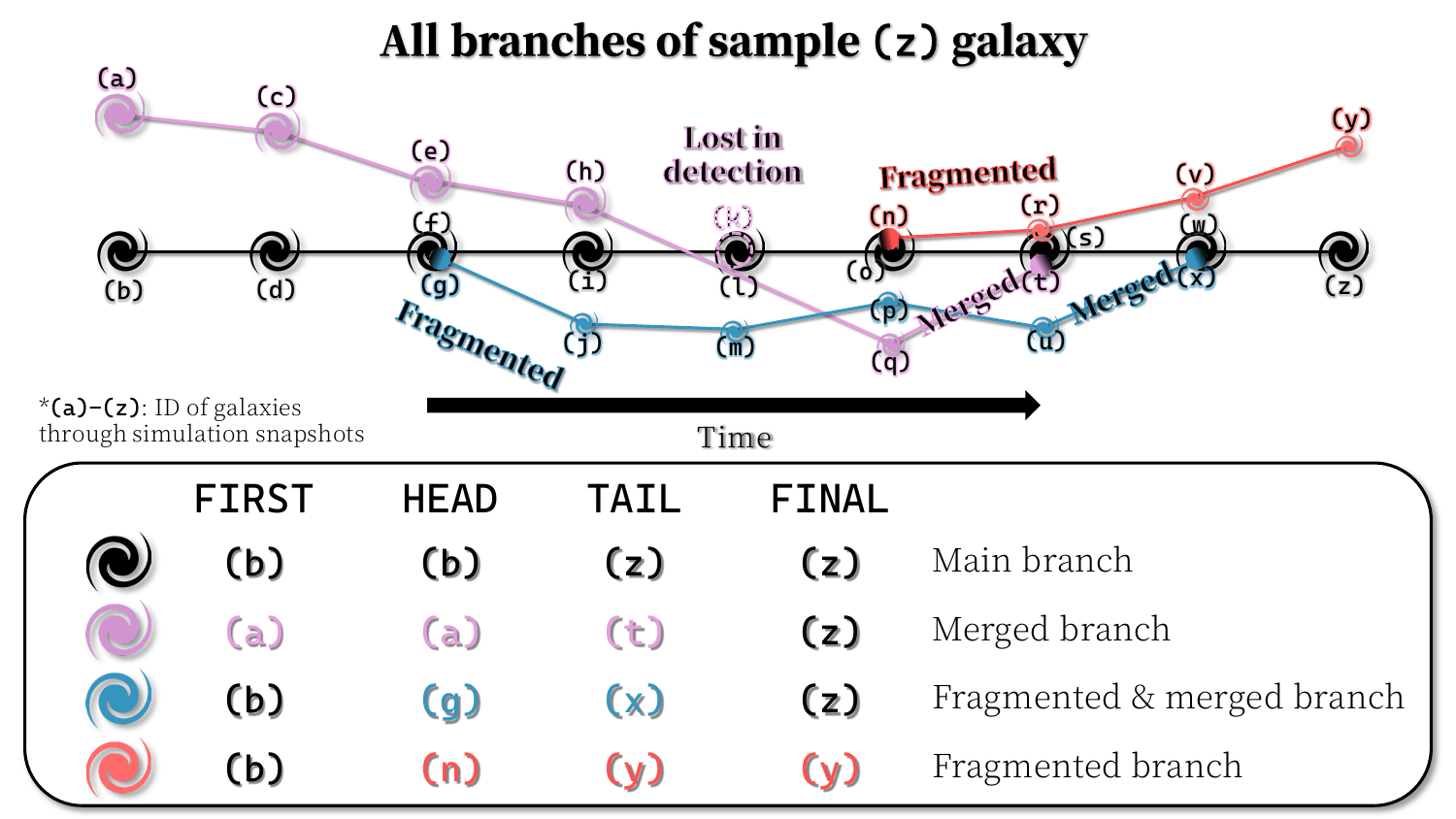}
    \caption{
         Schematic diagram of the merger tree.
         The spiral icons indicate all leaf galaxies connected to the target galaxy \gid{z} across different snapshots, marked with alphabetical IDs.
         Each line represents a connected branch based on the score (described in the text).
         The main branch (black) has the same IDs for \yfrom\ and \yhead, indicating a non-fragmented branch.
         Likewise, the same IDs for \ytail\ and \ylast\ mean that this branch is not merged into any other branch.
        }
    \label{fig02_scheme}
\end{figure*}
%

\subsection{Building merger trees}\label{sec:2method_3tree}

Since halo finders provide halo/galaxy catalogs only for individual snapshots, constructing a merger tree is essential for studying the time-series evolution of galaxies.
Importantly, the merger tree must properly track low-mass galaxies in very dense environments, which is particularly crucial in galaxy cluster studies.
The simplest and conventional method is to calculate the shared member particles between two adjacent snapshots and connect progenitors and descendants with the highest number of shared particles.
However, this method can lead to a ``broken tree'' issue (see e.g., \citealt{Srisawat2013}; more recently \citealt{Poole2017, Park2022, ChandroGomez2025}).
Commonly, when small galaxies closely pass through a larger galaxy, most particles could be falsely allocated to the larger one, or worse, a halo-finder temporarily fails to identify small structures embedded in a high-density region.
Studies with high-resolution simulations have more technical problems due to, ironically, their good output resolutions.
The standard parameter set of halo-finders often overpredicts highly fragmented substructures in high-resolution simulations, requiring manual removal or adjustment.

To overcome this issue, we hereby introduce our in-house merger tree algorithm.\footnote{\eadd{The source code is available on GitHub (\texttt{YoungTree}: \url{https://github.com/syj3514/YoungTree.git}) under a 3-Clause BSD License, and version 1.0.0 is archived in Zenodo \citep{YoungTree}}.}
To robustly measure the connectivity among galaxies across different snapshots, we measure the combined scores of 1) the bidirectional match score using shared particles, 2) the velocity offset of shared particles, and 3) the stellar mass differences.
We first assign a proximity-based weight in phase space to each member particle.
Consider two galaxies, $A_{i}$ and $B_{j}$, at snapshots $i$ and $j$, with $N_{A_{i}}$ and $N_{B_{j}}$ member particles, respectively.
The weight ($\tilde{w}_{k|A_{i}}$) of the $k$-th member particle of the $A_{i}$ is defined as:
\begin{equation}
    \tilde{w}_{k|A_{i}} \equiv 
    \frac{m_k}{
    \sqrt{
    \left(\frac{\lVert \bm{r}_k-\bm{r}_{A_{i}} \rVert}{\sigma_{r,A_{i}}}\right)^2
    +
    \left(\frac{\lVert \bm{v}_k-\bm{v}_{A_{i}} \rVert}{\sigma_{v,A_{i}}}\right)^2
    }},
    \label{eq2}
\end{equation}
where $m_k$ is the $k$-th particle mass, $\bm{r}_k$ and $\bm{v}_k$ are the position and velocity vectors of the $k$-th particle, and $\bm{r}_{A_{i}}$ and $\bm{v}_{A_{i}}$ are the position and velocity of the host $A_{i}$.
$\sigma_{r, A_{i}}$ and $\sigma_{v, A_{i}}$ are the standard deviations of distances and relative velocities of all members in $A_{i}$, respectively.
We then divide the weights by the sum of them, $w_{k|A_{i}} = \tilde{w}_{k|A_{i}} / \sum_{p=1}^{N_{A_{i}}} \tilde{w}_{p|A_{i}}$, which ensures that $\sum_{k=1}^{N_{A_{i}}} w_{k|A_{i}} = 1$.

Using the weights, we define the match rate as
\begin{equation}
    S_{A_{i}}\big(B_{j}\big) \equiv 
    \sum_{k\in A_{i} \cap B_{j}} w_{k|A_{i}},
\end{equation}
where ${A_{i} \cap B_{j}}$ denotes the shared particles between $A_{i}$ and $B_{j}$.
We define the bidirectional match score,
\begin{equation}
    q^{(1)}_{A_{i} , B_{j}} 
    = S_{A_{i}}\big(B_{j}\big) 
    + S_{B_{j}}\big(A_{i}\big).
\end{equation}
If $A_{i}$ and $B_{j}$ have identical members, $q^{(1)}_{A_{i} , B_{j}}=2$.

We also incorporate kinematic information.
For a single galaxy across different snapshots, the velocity of its shared particles relative to the bulk motion should not vary significantly.
We therefore define a relative velocity of the shared particles to the host as
\begin{equation}
    \Delta\bm{v}_{A_{i}}\big(B_{j}\big)
    \equiv \bm{v}_{A_{i} \cap B_{j}} 
    - \bm{v}_{A_{i}},
\end{equation}
where $\bm{v}_{A_{i} \cap B_{j}}$ is the mean velocity of the shared particles, and $\bm{v}_{A_{i}}$ is the bulk velocity of $A_{i}$.
$\bm{v}_{A_{i} \cap B_{j}}$ and $\bm{v}_{A_{i}}$ are weighted averages using the normalized weights, $w_{k|A_{i}}$.
We then define the second score of velocity offset,
\begin{equation}
    q^{(2)}_{A_{i} , B_{j}}
    = 1 - \frac{\big\lVert \Delta\bm{v}_{B_{j}}\big(A_{i}\big) - \Delta\bm{v}_{A_{i}}\big(B_{j}\big) \big\rVert}
               {\big\lVert \Delta\bm{v}_{B_{j}}\big(A_{i}\big) \big\rVert + \big\lVert \Delta\bm{v}_{A_{i}}\big(B_{j}\big) \big\rVert}.
\end{equation}
The score $q^{(2)}_{A_{i}, B_{j}}$ is unity for identical kinematics and zero for opposite.

As a final criterion, we also consider the stellar mass difference,
\begin{equation}
    q^{(3)}_{A_{i} , B_{j}}
    = \left( \frac{ \min(m_{A_{i}}, m_{B_{j}}) }{ \max(m_{A_{i}}, m_{B_{j}}) } \right)^C,
\end{equation}
where $m_{A_{i}}$ and $m_{B_{j}}$ are the stellar masses of $A_{i}$ and $B_{j}$, respectively, and $C$ is a tunable exponent.
If $A_{i}$ and $B_{j}$ have similar masses, $q^{(3)}_{A_{i} , B_{j}}$ is close to unity.
Finally, we calculate the combined score as
\begin{equation}
    q_{A_{i} , B_{j}}
    = q^{(1)}_{A_{i} , B_{j}} + q^{(2)}_{A_{i} , B_{j}} + q^{(3)}_{A_{i} , B_{j}},
\end{equation}
for all galaxy pairs if they have at least one shared member particle.

Furthermore, since small galaxies are often missed in detection, we evaluate the score mentioned above not only between adjacent snapshots, but also across up to four snapshots (i.e., roughly $\pm 60\,\mathrm{Myr}$) before and after.
After scoring, all galaxies have their progenitor and descendant candidates in previous and subsequent snapshots.
If two galaxies mutually point at each other, we directly connect them into a single branch.
Otherwise, we connect two branches, but instead record them as merger or fragmented events.
As a result, the merger tree contains branches classified as main, merged, or fragmented.
We formalize this classification by tagging the \yfrom, \yhead, \ytail, and \ylast\ parameters:
\begin{itemize}
    \item \yfrom: first leaf ID of the tree (same \yfrom\ for the same origin),
    \item \yhead: first leaf ID of the independent branch (same \yhead\ for the same branch),
    \item \ytail: last leaf ID of the independent branch (same \ytail\ for the same branch),
    \item \ylast: final leaf ID of the tree (same \ylast\ for the same galaxy at the final snapshot).
\end{itemize}
From these parameters, we define a ``branch'' as a sequence of leaves sharing the same \yhead\ and \ytail, while a ``tree'' is defined as the leaves connected with the same \yfrom\ or \ylast.

We present the schematic diagram of an example tree of the target galaxy \gid{z} in Figure~\ref{fig02_scheme}.
The main branch (black) represents a simple case with no fragmentation (\yfrom\,=\,\yhead\,=\,\gid{b}) or merger (\ytail\,=\,\ylast\,=\,\gid{z}).
The purple line is a merged branch.
Initially, it has a different \yfrom\ (\gid{a}) from the main branch (black), but it merged later (\gid{t} to \gid{s}), resulting in a different \ytail\ (\gid{t}) from \ylast\ (\gid{z}).
In addition, the branch is not broken even when \gid{k} is not detected by \HM, because, as we mentioned, our tree algorithm checks the progenitor/descendant candidates in multiple snapshots.
The red and blue lines are fragmented branches from the main branch.
In most cases, they are star-forming clumps that are misidentified as individual galaxies by \HM.
Our tree algorithm can capture these structures by checking \yfrom\ and \yhead.

Our method builds on the standard technique of connecting galaxies via shared particles \citep{Srisawat2013} by incorporating kinematic and mass information, conceptually similar to other modern algorithms that use velocity information to improve accuracy \citep{Behroozi2013bCTREE, Elahi2019bTreeFrog}.
In a galaxy cluster, there are numerous galaxy-galaxy interactions or mergers with the BCG.
The key strength of our tree is its stability in connecting small galaxies in crowded environments.
We now robustly identify satellite galaxies that are still surviving or already disrupted based on our tree, as will be shown in Section~\ref{sec:2method_4satellite}.

%
\begin{figure}[tb!]
    \centering
    \includegraphics[width=0.40\textwidth]{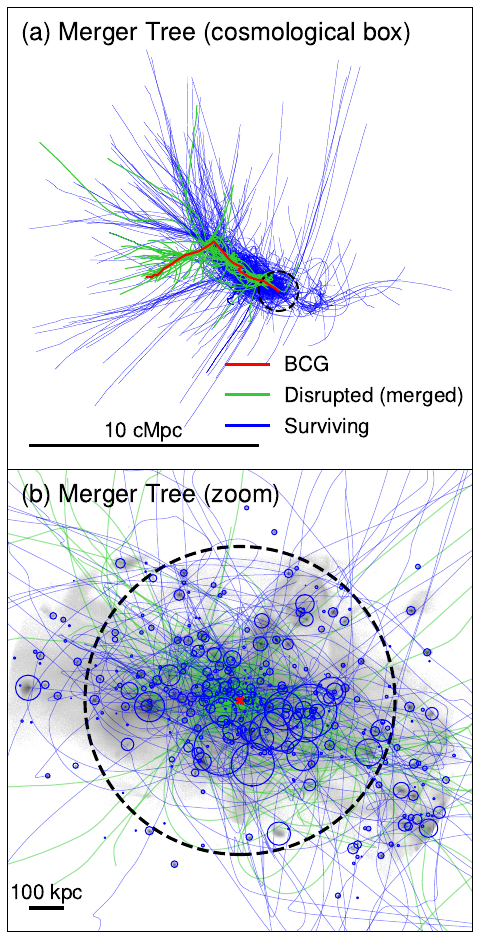}
    \caption{
         Orbital tracks of galaxies derived from the merger tree in the \NC\ cluster.
         Panel~(a) shows the comoving-scale tracks in the cosmological box, while panel~(b) shows the physical-scale tracks centered on the main cluster at $z=0.79$ overlaid on a stellar density map (grayscale).
         The black dashed circles mark the \rtwo\ of the main cluster halo.
         The red line in panel~(a) traces the BCG, and the red marker in panel~(b) indicates its final position.
         The green lines show satellites that have completely merged into the BCG.
         The blue circles indicate $R_{\rm 90}$, the radius enclosing 90\% of its stellar mass, and the blue lines show orbits of surviving satellites.
         Unlike panel~(a), we only present the tracks of surviving satellites more massive than $10^9\,\msol$ in panel~(b), for clarity.
        }
    \label{fig03_tree}
\end{figure}
%

\subsection{Satellite galaxies}\label{sec:2method_4satellite}
We identify the BCG from the galaxy catalogs as the most massive galaxy located at the center of the primary cluster halo.
For satellite galaxies, we first identify all satellite galaxies within $1.5\times\rtwo$, adopting a generous range.
We exclude satellites that have the same origin (\yfrom) as the BCG merger tree, which are likely fragmented structures of the BCG.
We further check the merger tree of the remaining satellites and exclude those originating from other satellites, because they are transient fragments of real satellites.
This selection process yields a final sample of 403 galaxies, which we define as ``surviving'' satellites.\footnote{Note that we do not apply a simple mass cut for sampling but select galaxies that have a persistent merger tree across time (e.g., 441 progenitors for each satellite on average), which ensures more robust identification by excluding transient structures.}
In Figure~\ref{fig03_tree}, we show their orbital tracks in blue lines, but only for galaxies more massive than $10^9\,\msol$ in panel~(b).

Since BCGs experience numerous merger events, the contribution of merged galaxies to the ICL is likely non-negligible \citep{Murante2007, Brown2024, Mayes2025, Joo2025}, although the relative contributions of merger and stripping depend on their definition \citep[cf.][]{Contini2018}.
To investigate this contribution, we also identify galaxies that have already merged and been disrupted.
We select all galaxies that have the same destination (\ylast) as the BCG tree but different origins (\yfrom).
We then apply two cleaning steps.
First, we exclude structures fragmented from the BCG if the maximum distance from the BCG is less than $1.5\times\rtwo$.
Second, we exclude fragments of surviving satellites.
This complex case occurs when any of the progenitors of a disrupted candidate is temporarily linked to the branches of surviving satellites.
If the origin (\yfrom) of this candidate is the same as that of any surviving satellites, we regard it as a fragment and exclude it.
After this, 67 galaxies are identified as ``disrupted (merged)'' satellites, and their tracks are shown in Figure~\ref{fig03_tree} (green).

\section{Classification of BCG+ICL stars}\label{sec:3class}

\begin{table}[b]
  \centering
  \caption{Classification of \bcgicl\ subsamples.}
  \label{tab:bcgicl_classes}
  \begin{tabular}{ll}
    \toprule
    Component & Definition \\
    \midrule
    \iclA & stripped from surviving satellites \\
    \iclB & stripped and disintegrated from disrupted satellites \\
    \iclC & born in the BCG \\
    \iclD & \parbox[t]{0.62\linewidth}{%
      from satellites but unbound before infall by group environments or neighboring galaxies%
    } \\
    \bottomrule
  \end{tabular}
\end{table}

%
\begin{figure}[htb!]
    \centering
    \includegraphics[width=0.40\textwidth]{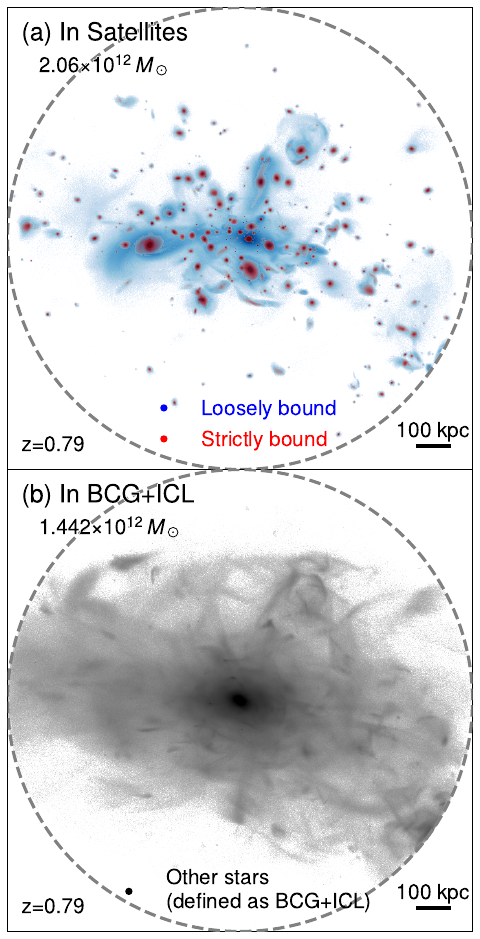}
    \caption{
         Illustration of the definition of the ICL within $1.5\times\rtwo$.
         Panel~(a) shows all stars bound to satellite galaxies at $z=0.79$, while panel~(b) presents the density map of the final \bcgicl\ component, after excluding the satellite stars shown in panel~(a).
         As described in the main text, the tight (red) and loose (blue) member stars are identified and follow the motion of their host satellite galaxy.
        }
    \label{fig04_ICL}
\end{figure}
%

The ICL is conventionally defined as the diffuse stellar component bound to the cluster potential but unbound from individual satellite galaxies.
However, it is not trivial to determine the true stellar membership of an individual galaxy.
For example, the stars in the galactic outskirts are barely bound to the host galaxy and are often not identified as members and instead classified as ICL.
However, when they are still kinematically aligned to the host and follow the motion of the host, this method potentially leads to an overestimation of the ICL.
Alternatively, direct binding-energy calculations combined with kinetic information are more robust \citep{ContrerasSantos2024, Joo2025}, but uncertainties in center finding and, more practically, the computational cost are prohibitive for high-resolution simulations \citep{Rudick2009, Puchwein2010}.

\eadd{Furthermore, decomposing the BCG and ICL components is one of the main challenges \citep{JimenezTeja2016}.
It is also physically ambiguous whether the BCG and ICL are truly separated populations or the ICL is just an extended feature of BCGs \citep{Donzelli2011, Mihos2016, Pillepich2018, Montes2019}, apart from the technical difficulty in the separation \citep{Contini2022}.
In observations, two- (or three-) component fitting and fixed-aperture masking are widely used to distinguish those \citep{Gonzalez2005, Annunziatella2016, Kravtsov2018, Spavone2022, Ahad2023, Bellhouse2025}, acknowledging the overlap of populations.
On the contrary, most theoretical studies adopt their own criteria; conventionally, particle membership from halo finder results \citep{Dolag2010, Pillepich2018, AlonsoAsensio2020, Yoo2022, MontenegroTaborda2023, Brown2024, Kimmig2025}.
This method has the drawback of missing stars in the outskirts (i.e., stellar halos), which are supposed to contribute significantly to the ICL because of their weak boundness \citep{Mayes2025}.
Also, similar to observational limits, a halo finder relying on spatial distribution cannot distinguish true galaxy members from ICL stars temporarily overlapping with galaxies.
To make matters worse, since most halo finders rely on single-snapshot data, stellar membership---and even galaxy detection itself---can vary strongly from snapshot to snapshot \citep{Behroozi2013bCTREE, Srisawat2013}.}

Considering these, to isolate the \bcgicl\ component from all stellar populations in the cluster, we employ the time-series membership approach, utilizing the merger tree.
The membership from \HM, at the very least, indicates that the stars are temporarily in the same spatial occupation of the galaxy.
We evaluate their kinematic alignment to the host (i.e., continue to follow the host's motion) by checking whether they consistently appear in the membership.
In particular, we classify the member stars of the surviving satellites within a $500\,\mathrm{Myr}$ time window before the final snapshot.
During this time, stars identified as members of the satellite galaxy in every post-birth snapshot are defined as ``tight,'' while those present in more than 10\% of their lifetime as ``loose.''
Although this threshold is somewhat arbitrary, it approximates binding without explicit potential calculations and is more robust than single-snapshot assignments.

Figure~\ref{fig04_ICL}(a) presents the resulting tight (red) and loose (blue) stellar members of surviving satellites.
In this work, to ensure a conservative definition of the \bcgicl\ components, we include the loose members in the satellite membership, which can be interpreted as the upper limit of the bound stars.\footnote{The mass fraction of the loose members in the total satellite membership is $\sim24\%$, a proxy for the uncertainty in our methodology.}
By subtracting these satellite members from the total stellar population, we define our final sample, \bcgicl, shown in Figure~\ref{fig04_ICL}(b).
The mass fraction of \bcgicl\ ($f_{\rm BCG+ICL}$) is $\sim41\%$, which is broadly consistent with other studies \citep{Pillepich2018, Brough2024, Kluge2025, Ellien2025, Mayes2025}.
As previous studies generally predict an increasing ICL fraction toward $z=0$ but no significant change for $z>0.5$ \citep{Burke2015, Joo2023, Contini2024a}, this provides another benchmark for studies of the ICL fraction's evolution at intermediate and high redshifts.

The final subsamples of \bcgicl\ are classified into four components based on their origin: \iclA, \iclB, \iclC, and \iclD.
The definitions of each component are summarized in Table~\ref{tab:bcgicl_classes}, and the detailed procedures are described in the following subsections.

\subsection{\bcgicl\ from surviving satellites}\label{sec:3class:1surv}

%
\begin{figure*}[htb!]
    \centering
    \includegraphics[width=0.95\textwidth]{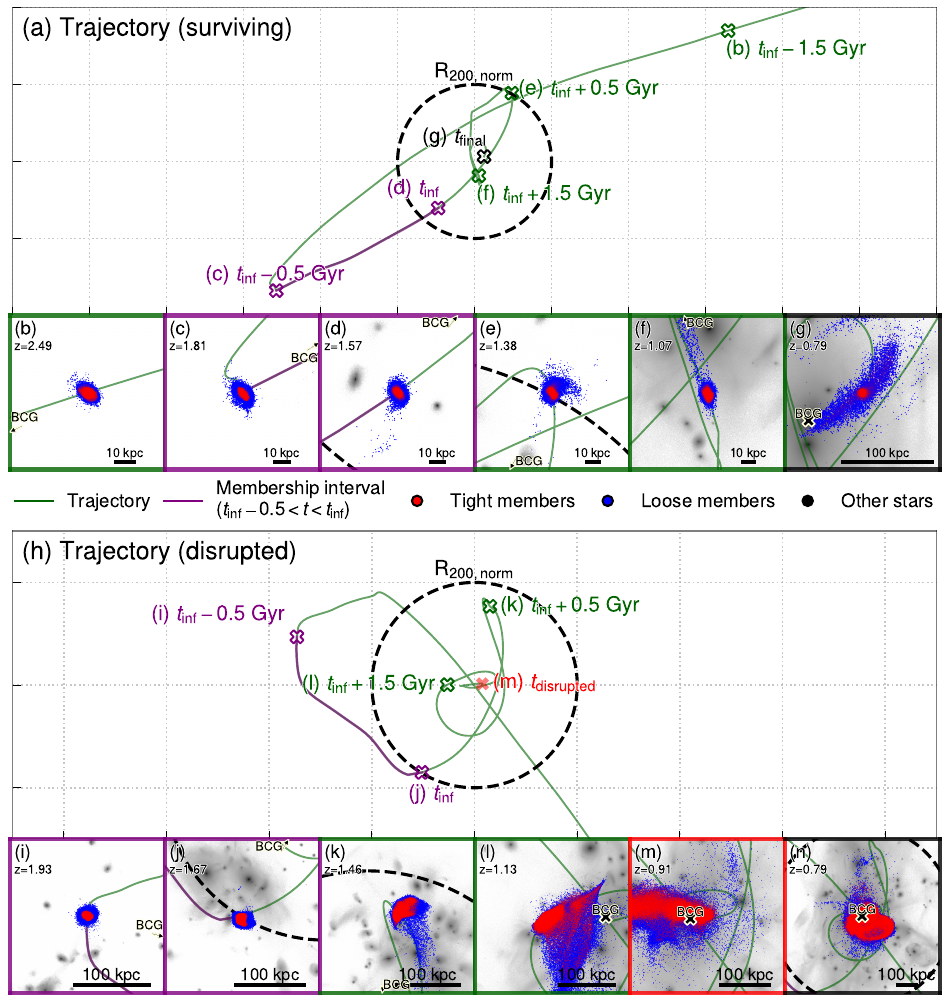}
    \caption{
         Top~(a)-(g): example of the infall member stars carried by a target surviving satellite galaxy.
         Panel~(a) shows the overall projected smoothed trajectory (green) of the target satellite.
         Each position is normalized by the \rtwo\ of the main cluster progenitors.
         The purple line highlights the time interval ($500\,\mathrm{Myr}$ before infall) used to identify member stars.
         Panels~(b)-(g) present zoom-in views of the target satellite at each stage, showing the tight (red) and loose (blue) member stars along with other stars (grayscale).
         We also indicate the direction of the BCG with arrows and markers.
         Bottom~(h)-(n): example of infall stars associated with a target disrupted galaxy.
         Similar to the top panels, panel~(h) shows the overall projected trajectory, and the other panels present zoom-in views.
         Since the target does not survive to the final snapshot, we include $t_{\rm disrupted}$ (red), the epoch when the satellite is last detected.
        }
    \label{fig05_example}
\end{figure*}
%

For each surviving satellite, we trace its full progenitor history using the merger tree (sharing the same \yfrom\ or \ylast) and then extract all member particles of any progenitors within a window from $500\,\mathrm{Myr}$ before infall to the first infall (entering \rtwo\ of the cluster, \tinf).
As in Section~\ref{sec:3class}, we classify stars as tight ($>90\%$), loose (10--90\%), or non-members ($<10\%$) by their occurrence across this window; for stars born within it, by the fraction of their lifetime.
We define the ``\iclA'' components as stars that are initially tight or loose members of surviving satellites but are later stripped into the \bcgicl.
Inspecting orbits, we find that a small fraction of loose members become unbound even before cluster infall, which arises from uncertainties in the binding criterion.
We therefore reclassify them as \iclD\ instead of \iclA\ components.
The \iclD\ components are discussed in Section~\ref{sec:3class:4pre}.

In the top panels of Figure~\ref{fig05_example}, we present the track and stellar distribution of a sample satellite.
Panel~(a) shows the satellite's main-branch track (green line), normalized by \rtwo\ of the cluster progenitor at each time, and other panels present zoom-in views exhibiting the \iclA\ stellar distribution (red and blue) over a grayscale background of other stars.
The membership remains well sustained before infall---panels~(b)--(c)---, whereas after infall the \iclA\ components are removed by the cluster tidal field, with the effect especially severe for the loose members---panels~(d)--(g).
For completeness, we also include stars formed within $500\,\mathrm{Myr}$ after \tinf\ in the satellite, although they are not shown in Figure~\ref{fig05_example} for clarity.

\subsection{\bcgicl\ from disrupted satellites}\label{sec:3class:2disr}

We identify disrupted (or merged) galaxies as those sharing the same final destination (\ylast) as the BCG but with a different origin (\yfrom), and determine their stellar membership using the same criteria as for the surviving satellites. 
The subset of these members later found in the \bcgicl\ is defined as the ``\iclB'' component, after reclassifying any pre-infall unbound stars as \iclD.

The bottom panels of Figure~\ref{fig05_example} display the track and distribution of the \iclB\ components for a sample galaxy, in the same format as the top panels.
Since the target is disrupted, we highlight in red the snapshot when this satellite was last detected ($t_\textrm{disrupted}$).
As with surviving satellites, the \iclB\ components remain well bound prior to infall.
However, after infall, they are strongly stripped, even more severely than for surviving satellites.
At the final snapshot (Figure~\ref{fig05_example}(n)), both the tight (red) and loose (blue) members are highly scattered along their orbital path, although the loose members show a more extended distribution.

\subsection{\bcgicl\ born in BCG}\label{sec:3class:3bcg}

Since the BCG has resided at the center of the overdensity, it likely formed a substantial number of in-situ stars through continuous gas accretion, in addition to those contributed by direct mergers \citep{Bonaventura2017, MontenegroTaborda2023, Mayes2025}.
To identify these in-situ stars, we extract the full merger tree of the BCG (same \yfrom), including its fragmented structures---mostly internal star-forming clumps---.
We classify stars associated with the extracted BCG tree based on the fraction of their lifetime spent as members, defining them as tight if the fraction is $>90\%$ and loose if it is $\ge10\%$.
These stars are collectively referred to as the ``\iclC'' components.

\subsection{\bcgicl\ from preprocessing}\label{sec:3class:4pre}

%
\begin{figure}[tb!]
    \centering
    \includegraphics[width=0.45\textwidth]{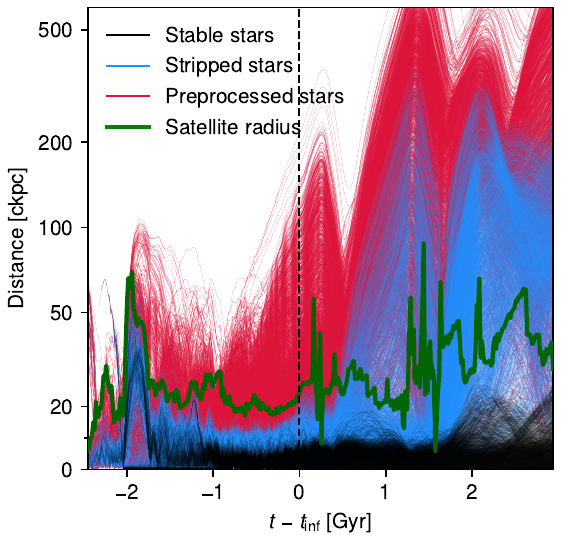}
    \caption{
         Example of stellar trajectories in a target satellite galaxy.
         The $x$-axis shows the time relative to infall ($t_{\rm inf}$; black dashed vertical line), and the $y$-axis is the comoving distance of stars from the satellite center.
         The black lines denote ``stable'' stars that remain tightly bound to the satellite throughout the simulation.
         The blue lines indicate ``stripped'' stars that are initially bound, but stripped after infall.
         The red lines represent ``preprocessed'' stars that are already unstable or unbound well before infall.
         The satellite radius is also shown at each time as a guideline for boundness (green).
        }
    \label{fig06_preprocess}
\end{figure}
%

Meanwhile, the environmental effects below the cluster scale or before the cluster infall (i.e., preprocessing) are now actively studied, mainly on the theoretical side.
For instance, \citet{Contini2024b} revealed the non-negligible contribution of preprocessing to the diffuse light fractions of low-mass ($M_{\rm halo}<10^{13}\,\msol$) halos.
Conventionally in the ICL field, ``preprocess'' refers to the ICL stars that initially reside in their host satellites but become unbound before cluster infall \citep{Fujita2004, Han2018, Jung2018, Contini2021, Contini2024a, JimenezTeja2025}.
However, in this work, we regard all processes that unbind previously bound stars before cluster infall (predominantly tidal stripping) as preprocessing, because tidal heating/stripping are largely self-similar across mass scales in the hierarchical structure formation.
To finalize the classification of the \iclD\ candidates mentioned in Sections~\ref{sec:3class:1surv} and \ref{sec:3class:2disr}, we evaluate their boundness by examining their orbital motions.
The kinematic signature of a \iclD\ component is generally either continued recession from its host or a sudden increase in apocenter distance well before infall.
In practice, stars showing clear evidence of prior unbinding are classified as ``\iclD'' components.
A visual justification is provided in Figure~\ref{fig06_preprocess}.

To justify and illustrate the classification of the \iclD\ components, we present the distances of member stars of a surviving satellite in Figure~\ref{fig06_preprocess}.
We display three populations: stable stars that consistently reside in the host to the end (black; satellite stars), stars tidally stripped after infall (blue; \iclA), and the \iclD\ stars that become unbound before infall (red).
The \iclD\ stars are evidently unbound and exhibit more extended orbits.
The main cause of preprocessing for this satellite is a major merger at $2\,\mathrm{Gyr}$ before \tinf, which produces the corresponding bump in its radius (green).

In summary, having classified the \bcgicl\ stars into four subsamples by their origin, we now proceed to investigate their properties.
We note that while a very recent work \citep{Mayes2025} applied a similar classification scheme, our detailed treatment of the \iclD\ components represents a key novelty.

\section{Demographics of the ICL in \NC}\label{sec:4demographics}

In this section, we present a demographic analysis of the \bcgicl\ stars by origin.
We note that the results represent a case study rather than a statistical sample, as our analysis is based on a single galaxy cluster.
Nevertheless, we emphasize that our results remain informative due to the unprecedented spatial and mass resolutions of stellar particles at the galaxy cluster scale.

%
\begin{figure*}[htb!]
    \centering
    \includegraphics[width=0.90\textwidth]{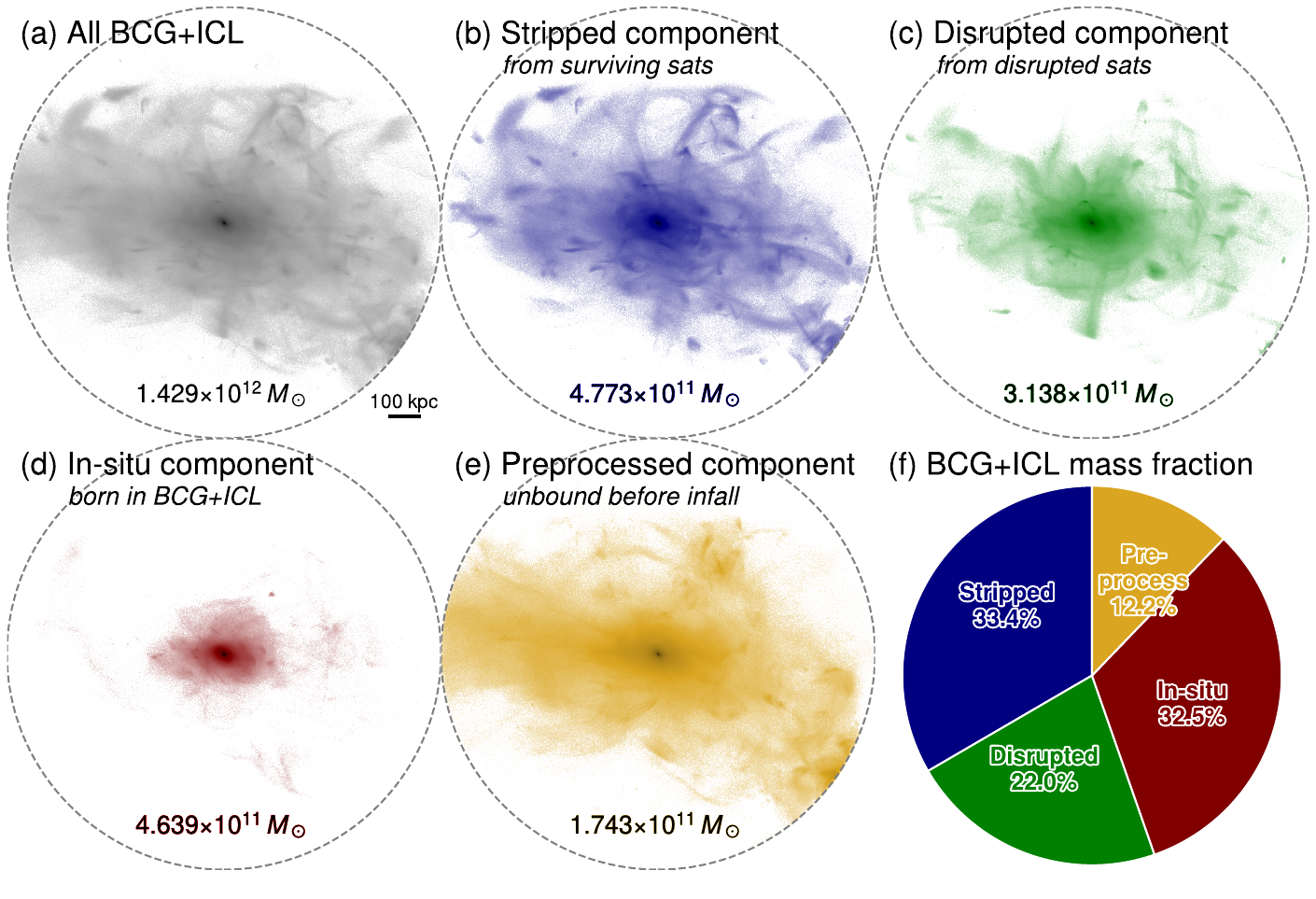}
    \caption{
         Stellar density maps of classified \bcgicl\ stars.
         Panel~(a) presents all \bcgicl\ stars within $1.5\,\rtwo$.
         Panels~(b)-(e) show the subsamples of the \bcgicl\ stars.
         Each subsample is defined as follows in Section~\ref{sec:3class} and displayed in a different color.
         Panel~(f) shows the mass fractions of the subsamples.
        }
    \label{fig07_maps}
\end{figure*}
%

First, we present the classification results.
Figure~\ref{fig07_maps} illustrates the stellar density maps of the classified populations.
Panel~(a) shows the overall \bcgicl\ sample, and panels~(b) to (e) display the subsamples of the \bcgicl: (b) those from surviving satellites, (c) from disrupted satellites, (d) born in-situ, and (e) from preprocessing.

The first noticeable feature is the tidal tail features stripped from surviving satellites (\iclA, blue) in Figure~\ref{fig07_maps}(b).
These stars are stripped inside the cluster and account for the largest fraction ($33.4\,\%$) of the \bcgicl\ components.
In terms of morphological features, they exhibit prominent, elongated tail-like structures, as their host satellites are still alive at the final snapshot.
We remind readers that these \iclA\ components were relatively tightly bound to the host satellite until infall, as we separate the \iclD\ populations that are already unbound before infall.
The stripped mass of $4.8\times10^{11}\,\msol$ remains small compared to the stellar mass of the surviving satellites ($2.1\times10^{12}\,\msol$; Figure~\ref{fig04_ICL}).

Figure~\ref{fig07_maps}(c) shows the stripped stars from satellites that {\em have been disrupted} (\iclB, green).
Most disrupted satellites are stripped in the tidal field of the cluster or merge into the BCG, rather than into other satellites in the cluster, due to their higher relative velocity \citep{Ghigna1998, Oh2018, Pearson2024}.
Moreover, these satellites typically enter the cluster earlier than the surviving ones, which gives them more time to reach dynamical relaxation, in addition to the smaller size of the cluster when they arrive.
As a result, this population exhibits a smoother and less extended distribution, compared to the \iclA\ ones.
Likewise, we can expect that the distribution of \iclA\ components would be relaxed toward lower redshift, as many currently surviving satellites complete mergers.

Figure~\ref{fig07_maps}(d) displays the \iclC\ components (red) that are predominantly born in the BCG, but also include a small minority of stars born in the ICL region.
While BCGs are generally thought to acquire the majority of their stellar mass from mergers \citep{AragonSalamanca1998, DeLucia2007}, we find that a comparable amount of stellar mass forms in-situ, which might be due to the current redshift ($z=0.79$), with an increased ex-situ fraction expected toward $z=0$.
We also note that there can be merger-induced star formation when satellites pass through the BCG \citep{Guo2011, Martin2017, Martin2021, ContrerasSantos2022, Aldas2025}.
This temporarily enhanced star formation can occur in both the BCG and its satellites; however, our method cannot separate them.
Nevertheless, despite this potential contamination, we observe that the distribution of the majority of the \iclC\ components is highly concentrated, suggesting that they have kinematics similar to that of the BCG.
Furthermore, we find that the majority of these stars formed deep within the BCG's potential well and do not migrate to the outskirts throughout their evolution.
As these in-situ stars occupy the central region, this population can provide a clue to the physical demarcation between the BCG and the ICL, a long-standing challenge in the field, as mentioned in Section~\ref{sec:1intro}.

The final population, \iclD\ components (yellow), is shown in Figure~\ref{fig07_maps}(e).
We find that the spatial distribution closely resembles the overall distribution of the \bcgicl\ stars.
As the \iclD\ components have experienced dynamical heating at earlier times, they exhibit a smoother and more extended distribution than the \iclA\ components that have been tightly bound to host satellites until recently.
In other words, the kinematics of \iclD\ components are more strongly governed by the cluster potential than by the potential of their original host satellites.
Given that the ICL is well regarded as a prospective and luminous tracer of dark matter \citep{Yoo2024}, we anticipate that \iclD\ better follows the dark matter distribution than the other subsamples---Figure~\ref{fig07_maps} shows stellar maps only, and we return to this point below.

We also present the mass fractions of the subsamples in Figure~\ref{fig07_maps}(f), with the color scheme of the sub-components in panels~(b)-(e).
The most noticeable feature is that all channels contribute comparable fractions to the total \bcgicl.
The \iclD\ component, for instance, accounts for $\sim12\%$, which is a substantial fraction, given that preprocessing is generally regarded as a secondary channel for ICL formation.
The dominant origin of \bcgicl\ is stars from satellites (surviving or disrupted), which account for more than half ($\sim55\%$) of the total \bcgicl.
\citet{MontenegroTaborda2023, MontenegroTaborda2025} have similarly investigated the demographics of the \bcgicl\ using the TNG300 simulation, broadly consistent with our results.
However, the detailed numbers slightly differ from ours, such as a low ($\sim19\%$) in-situ fraction or a larger \textit{merger} fraction than the \textit{stripped} fraction from surviving satellites.
This gap would be narrowed at lower redshifts as continued accretion increases the ex-situ fraction, and some of our currently ($z=0.79$) surviving satellites will eventually merge with the BCG.
In terms of spatial extent, the outer parts of the \bcgicl\ are mainly composed of accreted stars from satellites.
This indicates that the ICL, apart from the BCG, is predominantly formed by stellar stripping.
Stars formed in situ comprise a third of the total \bcgicl, but they rarely migrate to the outskirts.
\eadd{\citet{ContrerasSantos2024} also exhibit similar results with us, using \textsc{Gadget-X} and \textsc{Gizmo-Simba} codes based on \textsc{The Three Hundred} catalogs.
The \textsc{Gadget-X} simulation indicated a low fraction of in-situ stars ($12.5\%$) and a dominant fraction ($75.5\%$) of ``post-stripped'' components.
This may appear different from our results. 
However, the apparent difference is mostly coming from the difference in definitions.
They define ICL as isolated from the BCG, which corresponds to the outer part of our \bcgicl.
}

%
\begin{figure}[tb!]
    \includegraphics[width=0.45\textwidth]{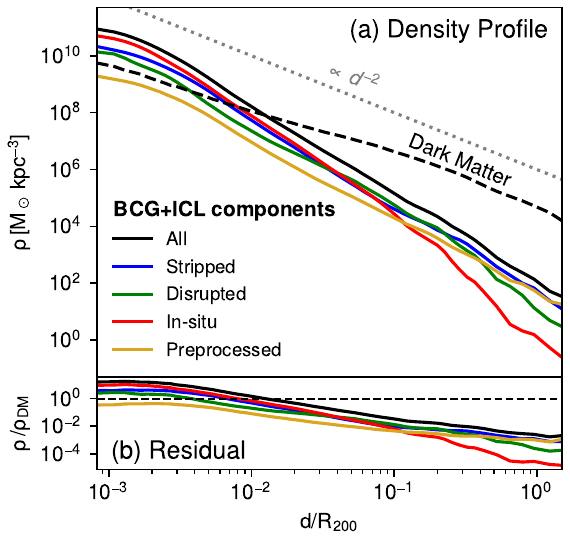}
    \caption{
         Density profiles of the \bcgicl\ stars and dark matter.
         Panel~(a) shows the density profile of dark matter (black dashed line) and all \bcgicl\ stars (black solid line).
         The subsamples of \bcgicl\ stars are shown in different colors as indicated in the legend.
         The gray dashed line indicates a slope of $-2$ for reference.
         Panel~(b) presents the residuals with respect to the dark matter density profile.
        }
    \label{fig08_denprof}
\end{figure}
%

To quantify the similarity of spatial distributions, we present the density profile of dark matter and stellar components in Figure~\ref{fig08_denprof}(a).
The $x$-axis is the radial distance from the cluster center, normalized by \rtwo, and the $y$-axis is the three-dimensional density of each shell.
The dark matter profile of the main cluster is shown in the black dashed line, while the profiles of stellar components are represented in solid lines, colored according to the legend.
We also exhibit the stellar density profile divided by the dark matter density in Figure~\ref{fig08_denprof}(b).
Again, the horizontal dashed line at unity in panel~(b) indicates the dark matter density for reference.

The dark matter profile has a much shallower slope than the stellar components, and it shows a slight break near $r_s\simeq0.15\,\rtwo$, where $r_s$ is the scale radius of the NFW (\citealt*{Navarro1997}) profile formula, exhibiting an indication of a weak core in the central region.
\eadd{The overall profiles of dark matter and the \bcgicl\ are consistent with those in other studies: dark matter profile being shallower than that of \bcgicl\ \citep{SampaioSantos2021, AlonsoAsensio2020, Diego2023, Butler2025}, and a roughly constant gradient in the ratio of the two density profiles \citep{AlonsoAsensio2020, Reina-Campos2022, Diego2023, Butler2025, Manuwal2025}.}
As expected, the \iclD\ population (yellow) has the shallowest slope in the density profile, making it the most similar to the dark matter profile of all the stellar components.
It is not necessarily expected that the stellar distribution should follow that of dark matter.
Most accreted stars inherit the biased phase-space of their parent satellites, whose orbits are reshaped by dynamical friction \citep{Chandrasekhar1943, Binney2008}.
In particular, recently formed stars originate in dissipative, shock-triggered star formation and are born on a biased subset of orbits, with the gas itself following trajectories distinct from the collisionless dark matter.
Hence, the closer similarity of the \iclD\ component to the dark matter can be understood as a mixing-time effect.
Stars unbound well before cluster infall have completed multiple orbital periods and experienced substantial phase-angle mixing, erasing coherent tidal features \citep{Helmi1999}.

\section{Stellar population study}\label{sec:5population}

%
\begin{figure}[htb!]
    \includegraphics[width=0.45\textwidth]{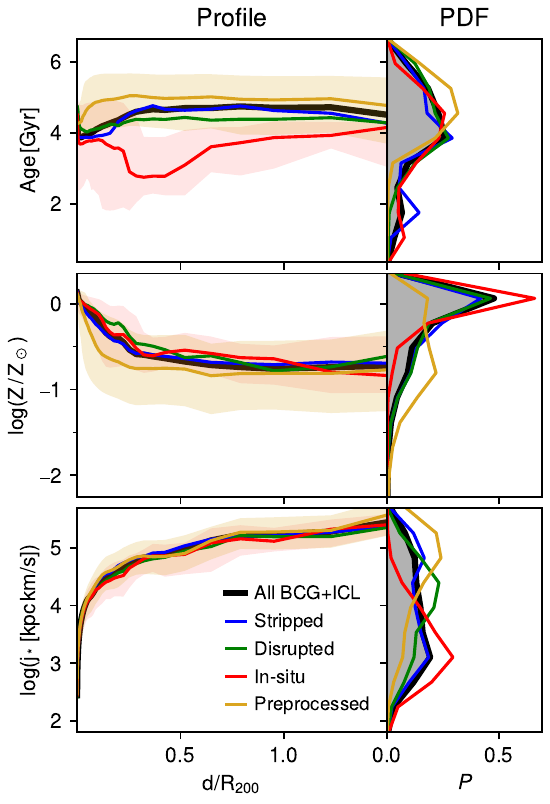}
    \caption{
         Physical properties of the \bcgicl\ stars.
         The left panel shows the radial profile, and the right panel shows the probability density function.
         From top to bottom, the properties shown are stellar age, metallicity, and specific angular momentum.
         All properties are mass-weighted.
         In the radial profiles, the 16th--84th percentile ranges of the \iclC\ (red) and \iclD\ components (yellow) are shown as shaded regions.
        }
    \label{fig09_props}
\end{figure}

\eadd{Thanks to advanced observational techniques, our understanding of the ICL is expanding beyond simple abundance measurements.
Many recent works have investigated detailed photometric properties.
\citet{Montes2021} found changes in the slopes of the surface brightness and color profiles in Abell 85 using Subaru Hyper Suprime-Cam, and more recently, \citet{Ellien2025} and \citet{Englert2025} also investigated the color profiles of the ICL of Abell 2390 and Abell 3667, respectively, finding that the profile can be varied by modeling methods or dynamical states of clusters.
However, the general stellar population of the ICL has not yet been systematically examined \citep{Montes2018, Contini2021}.}
Having analyzed the origins and spatial distributions of the \bcgicl\ subsamples, we now turn to their stellar population properties.
Figure~\ref{fig09_props} displays the analysis of the stellar age (top), metallicity (middle), and specific angular momentum (bottom).
The left (right) panel shows the mass-weighted radial profile (the probability density function).
The $y$-axis indicates the values of the corresponding properties.
The $x$-axis of the profile panels is the clustocentric distance normalized by \rtwo, and we also show the 16th--84th percentiles of the \iclC\ (red) and \iclD\ components (yellow) as shaded regions to highlight the most distinctive populations.

First, we examine the stellar age (first row).
At $z=0.79$, the age distribution peaks at $\sim4\,\mathrm{Gyr}$, which corresponds to $z\sim2.3$ or the cosmic noon \citep{Madau2014}.
As seen in the radial profile, the central region ($<0.3\,\rtwo$) exhibits younger populations.
The most noticeable difference among populations is that the \iclB\ (green) and \iclD\ (yellow) components are dominated by old populations, while the \iclC\ (red) and \iclA\ (blue) components are young.
Specifically, although all populations have similar age peaks near $\sim4\,\mathrm{Gyr}$, the \iclC\ and \iclA\ components show pronounced tails toward younger ages in the PDF.
This reflects recent or ongoing star formation inside the BCG and infalling satellites.
On the contrary, the \iclB\ and \iclD\ components rarely include young stars.
The \iclD\ components become unbound before the infall of satellites, and therefore have an inherent lower age limit set by the infall epoch of their host satellites.
Likewise, the \iclB\ components are naturally older than the \iclA\ components because their host galaxies stopped forming stars earlier than the host satellites of the \iclA\ components.
\eadd{This echoes the earlier study of \citet{Murante2004}, using a hydrodynamical cosmological simulation, which found that the diffuse components are older than other cluster stars on average.}

Next, we examine the metallicity (second row).
From the perspective of stellar evolution, younger stars are likely more metal-enriched due to the enrichment of recycled gas.
Consistent with this expectation, \iclC\ components exhibit a distinct distribution compared to other populations, showing a more skewed inclination towards higher metallicity.
In contrast, \iclD\ components display the opposite behavior, forming a predominantly metal-poor stellar population.
\eadd{Although the age or metallicity of ICL by different formation channels has hardly been studied}, the metallicity of the \bcgicl\ component in our sample is broadly consistent with previous studies \citep{Contini2019, Mayes2025}, exhibiting the drop of $\rm \log(Z/Z_\odot)$ toward the outskirts.
For the overall metallicity profile, the radial variation in metallicity is much more prominent than what can be inferred from the age difference.

We suggest that this difference in metallicity may provide a potential criterion for distinguishing the BCG and the ICL.
By our definition, \iclC\ components in our study represent the in-situ stellar population, which is mainly born inside the BCG.
In contrast, our \iclD\ components represent the accreted stars that are already free from the potential of individual galaxies, reflecting the mass growth driven by the surrounding environment.
This intrinsic difference in stellar populations can be traced by their metallicity, producing a potential demarcation of $Z\sim 0.5 \, Z_\odot$, where the two curves (red and yellow) intersect in the PDF panel, which roughly corresponds to a radius of $0.1-0.2\,\rtwo$ in the profile panel.

%
\begin{figure}[tb!]
    \centering
    \includegraphics[width=0.45\textwidth]{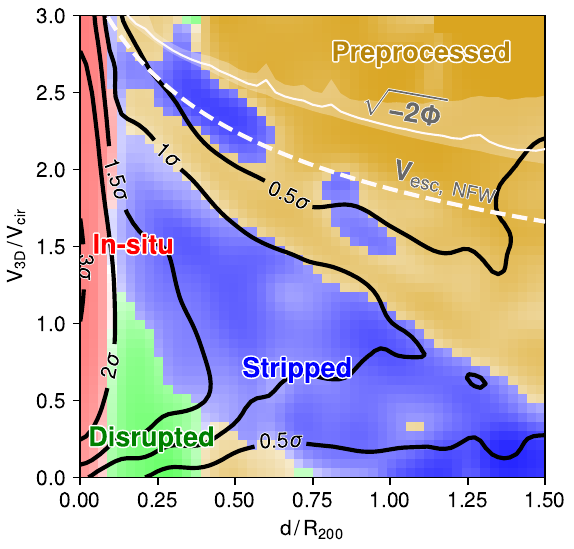}
    \caption{
         Dominant populations in the phase-space diagram.
         For each phase-space bin, we color the most dominant populations following the text shown.
         The opacity represents how dominant a population is in each bin; colors range from white (32\%) to vivid (100\%).
         The black contours indicate 0.5 to 3 $\sigma$ of the density distribution for all \bcgicl\ in phase-space.
         The white dashed line shows the escape velocity curve of an NFW ($c=6.5$) halo, while the white solid line (shaded region) represents the median (16th--84th) escape velocity directly derived from the gravitational potential.
        }
    \label{fig10_PSD}
\end{figure}
%

Lastly, we examine the specific angular momentum ($j_*$) of each population to investigate the kinematic differences.
The overall distributions, as shown in the PDF, are clearly distinct.
For example, the \iclD\ components exhibit higher $j_*$ values, peaking at $\sim10^5\,\rm{kpc\,km\,s^{-1}}$, whereas the \iclC\ components peak around $10^3\,\rm{kpc\,km\,s^{-1}}$.
However, this difference in $j_*$ is primarily a consequence of the different spatial distributions among populations, rather than an independent kinematic signature\, as shown in the radial profiles of $j_*$, which are nearly identical across all populations.
This suggests that the populations do not differ intrinsically in angular momentum at a given radius, but rather in their radial distributions.
Thus, $j_*$ does not provide an independent criterion for separating the \bcgicl\ components.

Nevertheless, we can use the kinematic information of the \bcgicl\ stars to qualitatively separate the populations.
In Figure~\ref{fig10_PSD}, we present the phase-space diagram, with each bin colored by its predominant component.
For example, the blue-colored regions indicate where the \iclA\ components (i.e., stripped stars from surviving satellites) dominate.
We note that the predominance does not necessarily mean that one component outnumbers the others at each bin.
We therefore adjust the opacity following the fraction of the dominant population.
In other words, the most white bins indicate the minimum fraction of $\sim32\%$, while the most vivid bins represent $100\%$.
The $x$-axis shows the clustocentric distance normalized by \rtwo, and the $y$-axis shows the three-dimensional velocity normalized by the maximum circular velocity ($\rm V_{cir}=825\,\textrm{km\,s}^{-1}$) of the cluster.
We also mark the $0.5 \textendash 3\,\sigma$ contours (black) of the overall \bcgicl\ components distribution and display the escape velocities derived from the NFW profile (white dashed line) and the gravitational potential (white solid line and shades).
In the central region within $\sim0.1\,\rtwo$, most stars are \iclC\ (red) regardless of their velocity.
The \iclB\ components (green) occupy the central ($d<0.4\,\rtwo$) region, but notably have lower velocities, typically less than $\rm V_{cir}$.
By contrast, the \iclA\ components (blue) show a more extended distribution toward the outskirts and greater inhomogeneity in velocity space.
The outermost region with high velocities and large distances is predominantly occupied by the \iclD\ components (yellow).
It is also noteworthy that most unbound \bcgicl\ stars (i.e., above the escape velocity) are likely to be \iclA\ and \iclD\ components, indicating a first-infall phase.
To summarize, a radius of $\sim0.1\,\rtwo$ can serve as an empirical demarcation separating the in-situ BCG stars from those originating in satellites, which can be further divided using velocity information.

%
\begin{figure*}[htb!]
    \centering
    \includegraphics[width=0.90\textwidth]{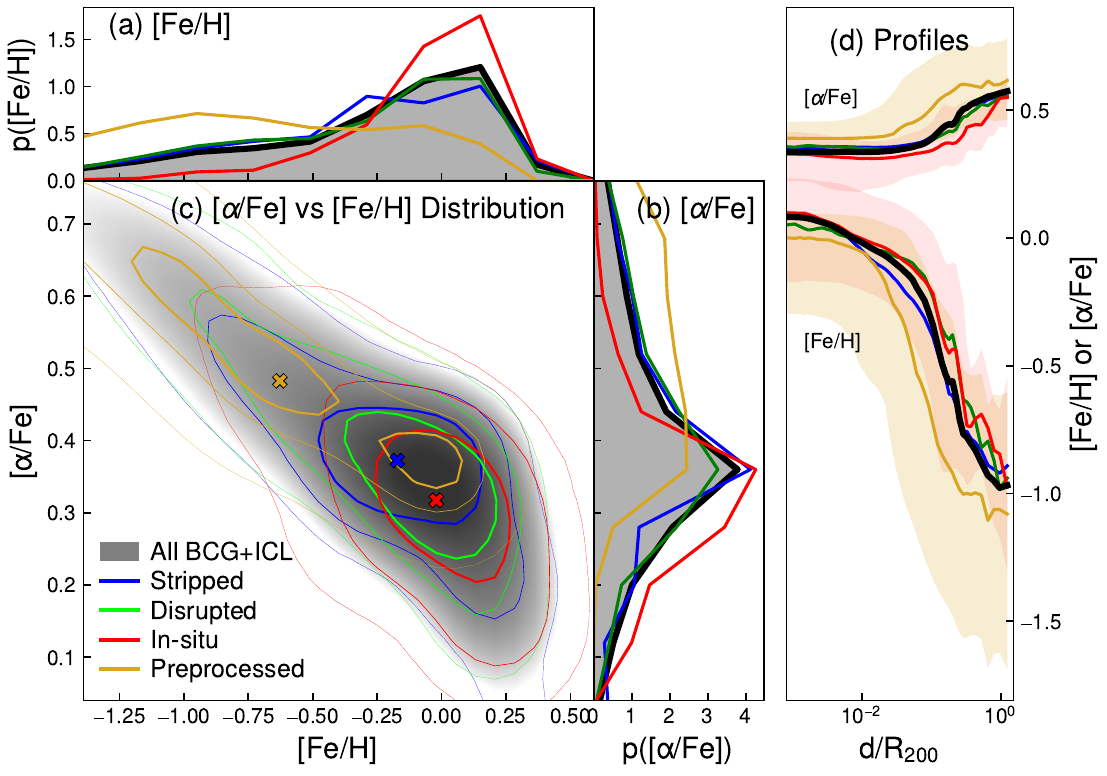}
    \caption{
         Distributions of \afe\ and \feh\ for the \bcgicl\ stars.
         Panel~(c) shows the \afe-\feh\ diagram for each subsample.
         The contours indicate 1, 1.5, and 2 $\sigma$ regions, while the shaded background shows the distribution of all \bcgicl\ stars.
         Panels~(a) and (b) display the probability density functions of \feh\ (left) and \afe\ (right).
         Panel~(d) presents the corresponding radial profiles.
         The \iclD\ component notably shows a distinctive distribution and profile, reflecting its different properties in the stellar population.
        }
    \label{fig11_alpha}
\end{figure*}
%

\NC\ computes the chemical evolution on-the-fly, which provides additional observable information.
We analyze the chemical abundance ratios of each subsample, specifically \afe\ and \feh.
\afe\ is calculated by averaging [O/Fe], [Mg/Fe], and [Si/Fe], all of which are directly available in \NC.
We use the solar chemical abundances from \citet{Asplund2009}.
Figure~\ref{fig11_alpha} presents the chemical distributions.
Panels~(a) and (b) show the PDFs of \feh\ and \afe, respectively, and panel~(d) exhibits radial profiles.
Panel~(c) displays the \afe--\feh\ plane, and different colors of contours indicate the \bcgicl\ subsamples following the legend.
We also show the distribution of all components as a grayscale background.
Again, the \iclD\ components occupy a distinct region compared to other components.
They show lower \feh\ and higher \afe, indicating older stellar ages \citep{Tinsley1979}.
Combining with the result that the \iclD\ components are old and metal-poor, as seen in Figure~\ref{fig09_props}, this suggests that the \iclD\ components formed in relatively sparse regions where gas recycling and chemical enrichment are inefficient.
The differences are still prominent in the radial profile.
In Figure~\ref{fig11_alpha}(d), unlike the results shown in Figure~\ref{fig09_props}, the \afe\ and \feh\ profiles of \iclD\ components show offsets from the mean profile (black) at all radii.
We also note that our result---super-solar at the center and $\textrm{\feh}<-0.5$ at the outskirts---is consistent with the observed profile \citep{Montes2018}.
In summary, our results highlight that a detailed analysis using various chemical elements can offer a more independent way to separate the stellar populations of the ICL.

\section{Relation between Orbital motions and stripping}\label{sec:6orbit}

%
\begin{figure*}[htb!]
    \centering
    \includegraphics[width=0.85\textwidth]{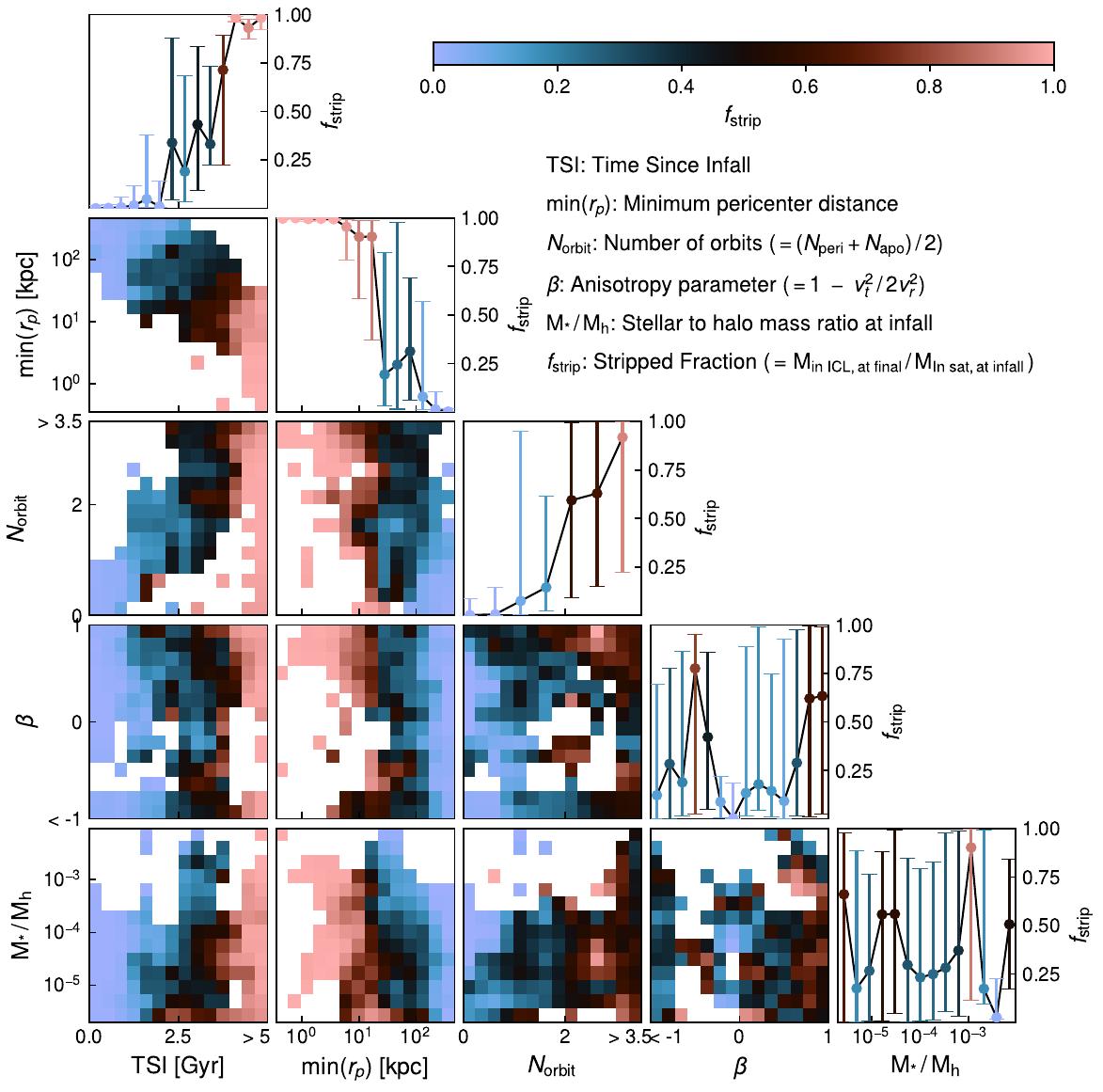}
    \caption{
         Relationships between orbital parameters and the stripped fraction ($f_{\rm strip}$).
         The parameters considered are the time since infall (TSI), the minimum pericenter distance ($\min(r_p)$), the number of orbits ($N_{\textrm{orbit}}$), the orbital anisotropy parameter ($\beta$), and the stellar-to-halo mass ratio at infall ($\rm M_* \, / \, M_h$), as described in the text.
         The diagonal panels show the relation between each parameter and $f_{\rm strip}$, including the 16th--84th percentile ranges.
         The off-diagonal panels present comparisons between parameter pairs, with colors indicating the value of $f_{\rm strip}$.
        }
    \label{fig12_orbit}
\end{figure*}
%

As shown in previous sections, most of the \bcgicl\ stars originate from satellite galaxies, both surviving and disrupted, whose spatial and velocity distributions differ.
The distinct properties among different stellar populations suggest that their formation is linked to the dynamical history of their host satellites.
In this section, we evaluate the relationship between the orbital parameters of the host satellites in the cluster and the efficiency of stellar stripping.

The measured orbital parameters are: 1) ``time since infall'' (TSI), the time since the first infall into the cluster; 2) $\min(r_p)$, the minimum pericenter distance; 3) $N_{\rm orbit}$, the number of orbits (i.e., half of the number of peri- and apocenters passages); 4) $\beta \equiv 1-v^2_t\,/\,(2v^2_r)$\footnote{Although this parameter is originally used to analyze the ensemble anisotropy of orbital systems, we extend this to the satellite galaxy orbiting the cluster for tracing the orbital shape.}, the characterized orbital anisotropy parameter \citep{An2006, Binney2008}, where $v_t$ and $v_r$ denote the tangential and radial velocities, respectively; and 5) $M_*\,/\, M_h$, the stellar-to-halo mass ratio measured at $t_{\rm inf}$, where $M_*$ is the stellar mass of the satellite galaxy at its infall and $M_h$ is the halo mass of the main cluster progenitor at the corresponding epoch.
We measure these parameters from $t_{\rm inf}$ to the final ($z=0.79$) snapshot.
To quantitatively assess the effect of the parameters, we define the stripped fraction ($f_{\rm strip}$) as the fraction of the satellite (both surviving and disrupted ones) member stars at infall that are found in the \bcgicl\ at the final snapshot.

We present these relationships in the corner plot shown in Figure~\ref{fig12_orbit}.
The diagonal panels show the one-dimensional relations between each orbital parameter and $f_{\rm strip}$, with the error bars indicating the 16th--84th percentile range.
The off-diagonal panels show $f_{\rm strip}$ as a function of two orbital parameters.
$f_{\rm strip}$ is color-coded from blue to red with increasing values.
The three parameters TSI, $\min(r_p)$, and $N_{\textrm{orbit}}$ have clear relations with $f_{\rm strip}$, consistent with previous studies of tidal stripping in cluster simulations \citep{Smith2016, Rhee2017, Han2018}.
$f_{\rm strip}$ increases as satellites spend more time in the cluster, approach closer to the cluster center, or complete more orbits, although the scatters are large.
By contrast, $\beta$ and $M_*\,/\,M_h$ show no clear dependencies.
The most significant relationship is between $\min(r_p)$ and TSI, indicating that satellites are more likely to lose their member stars if they fall into the cluster earlier and pass closer to the cluster center \citep[e.g.][]{Taylor2001}.
These two parameters consistently exhibit clear dependencies, as seen in the left two columns.
We also note that, for a fixed TSI, $M_* \, / \, M_h$ at infall shows a marginal negative trend with $f_{\rm strip}$, and satellites with higher $\beta$ (i.e., more radial orbits) are likely to be more stripped.
\eadd{These results largely agree with the previous study of \citet{Martin2024}, which studied the effects of orbital motions and found that satellites with more radial orbits and more orbital cycles exhibit larger stripped fractions in idealized environments.}

Taken together, this analysis suggests that the efficiency of stellar stripping from satellites, specifically into the \bcgicl\ components, is primarily driven by the time since infall and the depth of penetration into the cluster potential, rather than by their orbital shapes or the stellar-to-halo mass ratio at infall.
Satellites that have resided in the cluster longer, reached closer pericenter distances, and completed more orbital passages are more likely to lose their members into the \bcgicl.
This finding, which highlights the central role of orbital dynamics in shaping the origin of the ICL, is consistent with previous theoretical studies of satellite evolution \citep{Puchwein2010, Contini2018, Martin2024}.

\section{Caveats}\label{sec:7limit}
In this section, we discuss the current caveats of our study, including a limited sample, the difficulty of direct comparison with observations, and methodological arbitrariness.
These caveats are not trivial but have the value of emphasizing the complicated nature of the ICL and guiding us in future research directions.

\subsection{Limiting sample and dynamical state}\label{sec:7limit:1dyn}
The main limitation of our study is that it focuses on only one galaxy cluster and does not capture the cluster-to-cluster variation.
Since galaxy clusters have a wide range of mass and dynamical states \citep{Kravtsov2012}, it would be ideal to investigate a larger sample of clusters to ensure meaningful statistics.
In particular, the dynamical state of the cluster should impact the demographics and origins of the ICL because the formation and evolution of the ICL are naturally affected by the assembly history of the cluster itself \citep{Rudick2011, Mostoghiu2019, Contini2023}.
For example, \citet{Contini2023} used a semi-analytic model to evaluate halo concentration, which is directly connected to the dynamical state.
They found that clusters with higher halo concentrations or dynamically older have higher ICL fractions, especially for low-mass halos.
Also, \citet{Yoo2022} applied the weighted overlap coefficient method that quantifies the similarity of two-dimensional spatial distributions to the Galaxy Replacement Technique \citep{Chun2022} simulation, and found a stronger similarity in the relaxed cluster samples than dynamically younger ones.
\citet{Montes2019} also suggested that the baryonic distribution traced from X-rays does not match the total mass distribution in disturbed clusters, although they did not find a similar offset in the ICL distribution.
Observationally, the results regarding color and metallicity have not yet converged, which may also be related to dynamical states \citep[see references therein]{JimenezTeja2016, Contini2021}.
Our primary cluster has experienced recent mergers and is about to merge, indicating that it is a relatively unrelaxed system, which could lead to biased results.

\subsection{Current limitation of direct comparison with observations}\label{sec:7limit:2obs}
\NC\ has a high stellar mass resolution of $2\times10^4\,\msol$ and a spatial resolution of $\rm 68\,pc$ compared to other cluster-scale simulations---such as \textsc{Hydrangea} \citep{Bahe2017}, \textsc{RomulusC} \citep{Tremmel2019}, or \textsc{TNG-Cluster} \citep{Nelson2024}---, which allows faint structures and outskirts to be resolved. 
\citet{Han2025a_yomp}, the introductory work of \NC, showcased the mock radiative transfer images of galaxies in \NC, which resolve galactic structures in detail.
At the same time, they also introduced the mock observations of low-surface brightness features at $z=0.8$, and demonstrated that modern instruments can resolve faint structures.
However, \NC\ is just passing $z\sim\zing$, i.e., relatively high in redshift for practical observations, and a direct comparison with observations at this redshift is challenging.
For example, \citet{Ko2018} used a deep Hubble Space Telescope Wide Field Camera 3 (HST/WFC3) observation, but the surface brightness limit was still $\rm \sim 29 \, mag \, arcsec^{-2}$.
Moreover, \citet{Dacunha2025} pointed out that the data reduction process, such as background subtraction, should be treated carefully to obtain reliable ICL measurements.

In addition, we utilize the three-dimensional position and time series data, which are fully available only in simulations.
\eadd{Although we separate the \bcgicl\ stars into four different channels and show that they exhibit different stellar properties, this approach is technically infeasible to replicate with current observational techniques.
For example, our \iclD\ component has the most distinct properties, such as old age, low metallicity, and far distance from the cluster center in the phase-space.
However, in observations, the quantities to estimate those require high-resolution spectroscopy, which makes the direct comparison challenging.}

A fair and direct comparison requires constructing mock images under the same conditions as observations.
Many recent instruments, such as the James Webb Space Telescope (JWST), the Vera Rubin Observatory's Legacy Survey of Space and Time (Vera Rubin/LSST), and the Euclid mission, are capable of observing the low-surface brightness features, and surveys are planned to study the ICL \citep{Montes2022b, Martis2024, Martin2022, Brough2024, Englert2025, Kluge2025, Bellhouse2025}.
To meet these observational capabilities, we can perform mock observations by combining radiative transfer with on-the-fly dust calculation---which is provided by \NC---, enabling direct and fair comparisons between simulations and future surveys.
For instance, \citet{Byun2025} demonstrated that mock observations using on-the-fly dust calculations of \NC\ can reproduce the galaxy's morphology of the early Universe observed by JWST.
Furthermore, as resolved spectroscopic studies of low-surface brightness are now emerging \citep{Adami2016, Gu2020, Spavone2022},  we can even build mock integral field unit (IFU) data using the simulation.
While observing techniques gradually reach farther with time, simulations (\NC\ in this case) approach nearer to the local Universe.
The quantities derived in our study may not be directly verified through observations yet, but it is promising that both observations and simulations are nearing their front lines.

\section{Conclusions}\label{sec:8conclusion}

The main goal of this paper is to present the first results of an ICL population study with high resolution and to highlight the potential of next-generation cluster simulations for ICL studies.
We used the high-resolution zoom-in cluster simulation, \NC, at $z=0.79$ to examine the origins and stellar population properties of the \bcgicl.
To ensure a complete census of stellar particles, we developed a dedicated merger tree algorithm and tracked individual stars across closely spaced simulation outputs.

Our main results can be summarized as follows.
\begin{enumerate}
    \item We found that tracking the full orbital history of satellite galaxies is essential for robustly determining the boundness of stellar particles.
    The member stellar particles of satellites are tightly bound to their host before infall, but they are then significantly stripped as the satellites orbit within the cluster.
    However, the survival---or the cluster infall time---of host satellites strongly influences the final distribution of stripped stars.
    
    \item We derived the demographics of the \bcgicl\ based on their origin: \iclA\ (\bcgicl\ stripped from surviving satellites), \iclB\ (\bcgicl\ stripped and merged from disrupted satellites), \iclC\ (formed in situ within the BCG), and \iclD\ (unbound from their host satellites prior to the cluster infall).
    The majority of the \bcgicl\ components originate from satellite galaxies ($\sim60\%$), including surviving and disrupted ones.

    \item Our classification of the \iclD\ component reveals its distinctive properties and shows that it comprises a significant fraction ($\sim12\%$) of the total \bcgicl.
    They exhibit shallower density profiles than the other \bcgicl\ components and are more closely aligned with the dark matter profile.
    The stellar population of the \iclD\ components shows more dramatic differences in age, metallicity, and alpha abundance (e.g., older, metal-poor, and enhanced \afe), suggesting this population as a key fossil record of the cluster's dynamical history.
    We also introduced a clear division of the subsamples in the phase-space, which provides a useful diagnostic for the ICL composition.

    \item We investigated the effect of orbital parameters on the fraction of stripped stars ($f_{\rm strip}$).
    We found that the time since infall and the minimum pericenter passage distance are the primary drivers in determining $f_{\rm strip}$.
    The satellite galaxies that join the cluster earlier or reach closer to the BCG in their orbital motion are stripped more.
    Also, a larger number of orbital cycles ($N_{\rm orbit}$) increases the fraction of stripped stars.
    The shapes of orbits and the stellar-to-halo mass ratio at infall have only a marginal effect on $f_{\rm strip}$.
\end{enumerate}

While we focus on a single cluster and acknowledge the limitations in the investigation, these significant caveats highlight the intrinsic complexity of ICL studies rather than diminishing the significance of our results.
Nevertheless, the methodological novelty of our work lies in our detailed classification of stellar populations, tracking billions of particles across dense temporal outputs to separate the ICL into four distinct origins.
Our work suggests new directions for future ICL studies, particularly in understanding the role of preprocessing and in using detailed stellar properties to disentangle formation channels.

In closing, we emphasize the prospects of the upcoming low-redshift data from \NC.
The primary and secondary halos in \NC\ are currently in the process of merging ($z\sim\zing$), and the merged final cluster is expected to undergo violent mixing and relaxation.
This stage provides an opportunity to investigate the migration of stars among the BCG, satellites, and the ICL through the cluster merger.
Crucially, because we follow the same cluster across time, we can quantify the ICL's evolution within one halo rather than combining heterogeneous samples.
As the evolutionary trend of the ICL fraction in a single galaxy cluster has not yet converged, we can disentangle the methodological differences and cosmological dimming from the true evolution, which provides the prediction of the later growth of the ICL as a benchmark.
The forthcoming low-redshift data from \NC\ will provide a unique laboratory for bridging theoretical and observational studies of the ICL.


\section*{Acknowledgements}
This work was granted access to the HPC resources of KISTI under the allocations KSC-2021-CRE-0486, KSC-2022-CRE-0088, KSC-2022-CRE-0344, KSC-2022-CRE-0409, KSC-2023-CRE-0343, KSC-2024-CHA-0009, and KSC-2025-CRE-0031 and of GENCI under the allocations A0150414625 and A0180416216
The large data transfer was supported by KREONET, which is managed and operated by KISTI.
S.K.Y. acknowledges support from the Korean National Research Foundation (RS-2025-00514475 and RS-2022-NR070872).
J.R. acknowledges support from the Institut de Physique des deux infinis of Sorbonne Universit\'{e} and by the ANR grant ANR-19-CE31-0017 of the French Agence Nationale de la Recherche.
J.L. is supported by the National Research Foundation (NRF) of Korea grant funded by the Korea government (MSIT, RS-2021-NR061998).
J.L. also acknowledges the support of the NRF of Korea grant funded by the Korea government (MSIT, RS2022-NR068800).
G.M. acknowledges support from the UK STFC under grant ST/X000982/1.
J.K. is supported by KIAS individual Grant (KG039603) at the Korea Institute for Advanced Study.".
E.C. and S.J. acknowledge support from the Korean National Research Foundation (RS-2023-00241934)
T.K. was supported by the Yonsei Fellowship, funded by Lee Youn Jae.

\bibliography{reference}{}
\bibliographystyle{aasjournal}

\end{document}